\documentclass[journal=jpccck,manuscript=article,layout=twocolumn]{achemso} 
\usepackage{graphicx}
\usepackage{microtype}
\usepackage{tabularx}
\usepackage{amssymb}
\usepackage{pdfpages}
\usepackage{mathtools}
\usepackage{makecell}
\usepackage{amsmath}
\usepackage{setspace}
\usepackage{multirow}
\usepackage{pslatex}
\usepackage{placeins}
\usepackage{siunitx}
\usepackage{color}
\usepackage{wrapfig}
\usepackage{gensymb}
\usepackage{fixltx2e}
\usepackage{subfig}
\usepackage{array}
\usepackage[font={small,bf}]{caption}
\usepackage{float}
\usepackage{xr}
\let\oldmaketitle\maketitle
\let\maketitle\relax
\SectionNumbersOn
\allowdisplaybreaks

\author{Steven A. Hawks}
\affiliation{Lawrence Livermore National Laboratory, 7000 East Avenue, Livermore, CA 94550, United States}
\author{Ashwin Ramachandran}
\affiliation{Department of Aeronautics \& Astronautics, Stanford University, Stanford, CA 94305, United States}
\author{Slawomir Porada}
\affiliation{Wetsus, European Centre of Excellence for Sustainable Water Technology, Oostergoweg 9, 8911 MA, Leeuwarden, The Netherlands}
\altaffiliation{Soft Matter, Fluidics and Interfaces Group, Faculty of Science and Technology, University of Twente, Meander ME 314, 7500 AE Enschede, The Netherlands}
\author{Patrick G. Campbell}
\affiliation{Lawrence Livermore National Laboratory, 7000 East Avenue, Livermore, CA 94550, United States}
\author{Matthew E. Suss}
\affiliation{Faculty of Mechanical Engineering, Technion-Israel Institute of Technology, Haifa, Israel}
\author{P.M. Biesheuvel}
\affiliation{Wetsus, European Centre of Excellence for Sustainable Water Technology, Oostergoweg 9, 8911 MA, Leeuwarden, The Netherlands}
\author{Juan G. Santiago}
\affiliation{Department of Mechanical Engineering, Stanford University, Stanford, CA 94305, United States}
\author{Michael Stadermann}
\affiliation{Lawrence Livermore National Laboratory, 7000 East Avenue, Livermore, CA 94550, United States}
\email{stadermann2@llnl.gov}
\title{Performance Metrics for the Objective Assessment of Capacitive Deionization Systems}
\begin{document} 
	
\twocolumn[
\begin{@twocolumnfalse}
	\oldmaketitle
	\begin{abstract}
{\small In the growing field of capacitive deionization (CDI), a number of performance metrics have emerged to describe the desalination process. Unfortunately, the separation conditions under which these metrics are measured are often not specified, resulting in optimal performance at minimal removal. Here we outline a system of performance metrics and reporting conditions that resolves this issue. Our proposed system is based on volumetric energy consumption (Wh/m$^3$) and throughput productivity (L/h/m$^2$) reported for a specific average concentration reduction, water recovery, and feed salinity. To facilitate and rationalize comparisons between devices, materials, and operation modes, we propose a nominal standard testing condition of removing 5 mM from a 20 mM NaCl feed solution at 50\% water recovery for CDI research. Using this separation, we compare the desalination performance of a flow-through electrode (fte-CDI) cell and a flow between membrane (fb-MCDI) device, showing how significantly different systems can be compared in terms of generally desirable desalination characteristics.  In general, we find that performance analysis must be considered carefully so to not allow for ambiguous separation conditions or the maximization of one metric at the expense of another. Additionally, for context we discuss a number of important underlying performance indicators and cell characteristics that are not performance measures in and of themselves but can be examined to better understand differences in performance.}
	\end{abstract}
\end{@twocolumnfalse}
]

\section{Introduction}
\indent \indent Capacitive deionization (CDI) encompasses a group of water desalination technologies that uses cyclic charging and discharging of electrodes and the resulting electrosorption of ions to deplete or enrich a feed water stream.\cite{Biesheuvel2017a} Interest in CDI has grown substantially in recent years, with currently more than 150 publications annually reporting on the topic.\cite{Zhang2018} As often occurs in a growing device research field, a need has emerged for establishing guidelines for evaluating, reporting, and comparing device performance.\cite{Balducci2017,Khenkin2018,Gogotsi2011,Brousse2015,Editorial2015,Zuliani2015} At present, much of the work in CDI focuses on materials and methods to improve the efficiency and desalination capacity of the system, but many published studies are actually not comparable to one another. This is primarily because either certain critical performance parameters are not reported, or the desalination conditions explored are different in some substantial aspect. 

For desalination technologies such as reverse osmosis (RO) or electrodialysis (ED), performance metrics typically include energy consumption per volume of dilute effluent and a volume throughput parameter.\cite{Cohen-Tanugi2014,Park2017,Alvarado2014,Pan2017} This set of metrics is sufficient for comparison between removal methods only when the water recovery (WR), feed salinity, and total concentration reduction are the same.\cite{Park2017} In other words, energy and throughput metrics alone are insufficient to provide a comparable description of performance because these values depend critically on the separation conditions. Achieving comparable separation conditions in CDI is particularly difficult because the method usually removes a smaller, highly variable fraction of the ions. Since extrapolating results from one set of removal conditions to another is not straightforward, it is critical to control for the feed salinity, concentration reduction, and water recovery in a performance evaluation. 

To add to the complexity of the CDI performance landscape, the community has turned to a number of metrics to describe its desalination process. Some commonly used energy, throughput, and materials metrics in CDI are: the salt adsorption capacity (SAC, in mg/g of electrodes), average salt adsorption rate (ASAR, in mol/min/cm$^2$ or in mg/g/min),\cite{Zhao2013a,Kim2015d,Hemmatifar2016,Hawks2017,Wang2018} energy consumption per mole of salt removed (in kJ/mol kJ/g, or $kT$/ion),\cite{Zhao2012,Qu2016,Suss2015,Shang2017} energy normalized adsorbed salt (ENAS, in $\mu$mol/J),\cite{Hemmatifar2016,Hawks2017,Oyarzun2018} and specific energy consumption (SEC, in mg/J).\cite{Suss2015,Wang2018} While these metrics may be insightful for the processes relevant to CDI, it would be beneficial to settle upon a minimum-necessary set of measures that reflect economically relevant desalination performance. Additionally, it is important to distinguish between throughput and energy consumption, as these different desalination aspects often trade off with one another. Indeed, several studies have shown that SAC, ASAR, and ENAS, for example, are highly coupled with clear tradeoffs among them.\cite{Hemmatifar2016,Kim2015d,Hawks2017,Wang2018} For instance, adsorption rate can be increased at the cost of higher energy consumption or lower SAC.\cite{Hemmatifar2016,Kim2015d} Thus, both an energy and a throughput metric must be reported for a given separation in order to fully define performance.

With these considerations in mind, herein we present a framework for objectively comparing desalination performance in capacitive-based removal systems along with an example standardized deionization protocol for CDI research. We also discuss the relation between the proposed parameters and their typically-used traditional counterparts. Our goal here is to create a framework by which generic, economically relevant, desalination capability can be evaluated for a given separation with any CDI-type approach. To highlight how this framework can be implemented, we propose a protocol that can be easily followed for both large and small laboratory-scale devices with the intent of facilitating study-to-study comparisons in CDI research. Hence, for example, our procedure does not prescribe flow rates, charging methods, device size, electrode configuration, or even removal mechanism. For context, we also distinguish between performance ``metrics" and performance ``indicators", where metrics are meant to reflect generic desalination performance and indicators are used to better understand why a certain performance was achieved. Since understanding the reasons behind differences in performance is vitally important to technological progress, we discuss a number of performance indicators and highlight the essential cell characteristics that underlie many of these results.  
\section{Framework for Comparing Deionization Performance}
\begin{figure}[h]
	\centering
	\includegraphics[width=1\columnwidth]{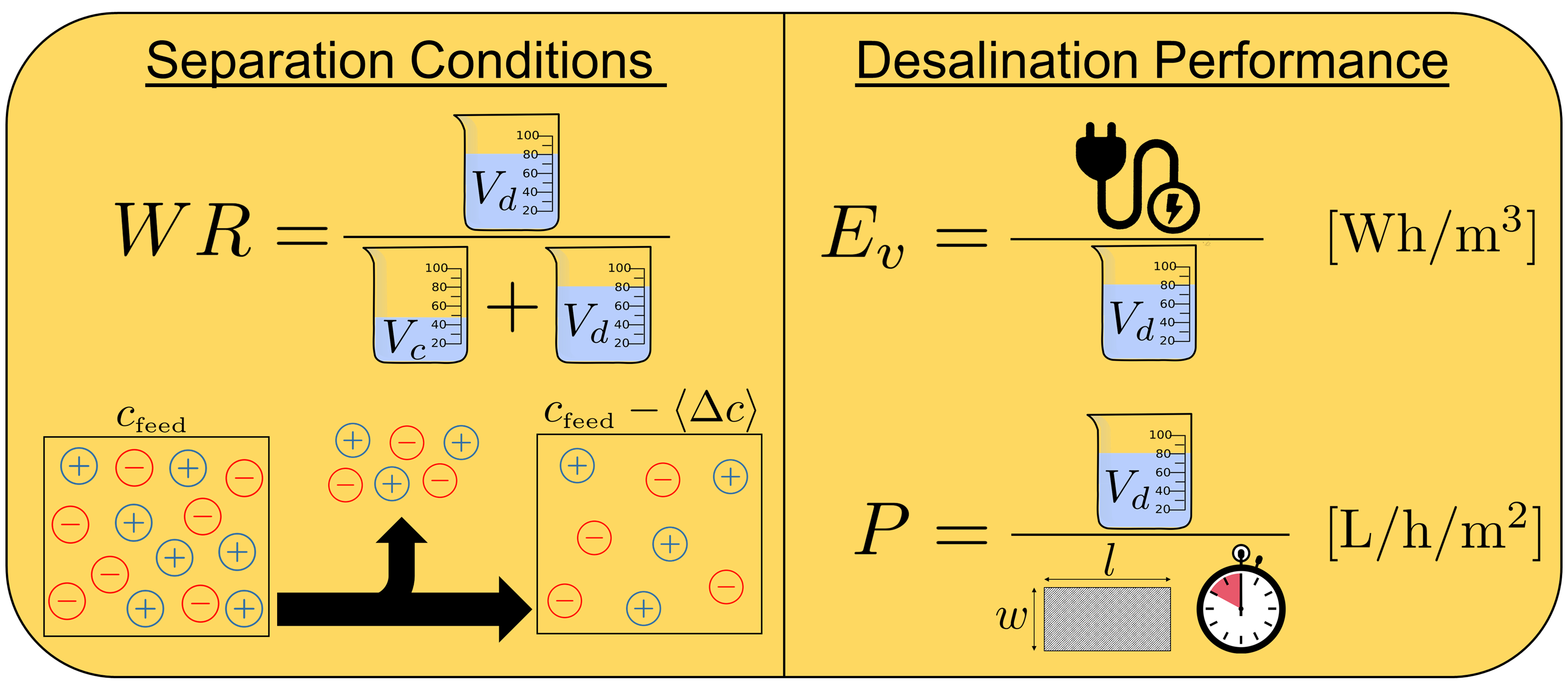}
	\caption{A pictogram illustrating how to objectively compare deionization performance. First, the separation conditions must be defined (water recovery, feed concentration, and concentration reduction), and then the desalination performance (energy consumption, throughput) can be compared. The tradeoff between energy consumption and throughput must be examined for equivalent separation conditions. The variables in the pictogram are defined below.}
	\label{Fig: pictogram}
\end{figure}

\indent \indent Our proposed approach to evaluating desalination performance of a CDI system is pictographically shown in Fig. \ref{Fig: pictogram}. First, and most importantly, a target separation must be defined, which includes the feed salt concentration ($c_{\text{\tiny feed}}$), the concentration reduction after treatment ($\left\langle \Delta c \right\rangle$), and the water recovery ($WR$). With these conditions defined, the desalination performance can be evaluated by examining the energy consumed per cubic meter of diluate, $E_v$, in Wh/m$^3$ and the throughput productivity, $P$, in L/h/m$^2$, where the area unit in $P$ corresponds to the projected face area of a typical device (i.e., the face area over which electrodes overlap, see Fig. S2 in the supporting information) or projected face area of the membranes in a flow-electrode cell.\cite{Cohen-Tanugi2014,Pan2017,Doornbusch2016} These performance metrics are defined quantitatively below for dynamic steady state (DSS) cycling but pertain to any operation mode (e.g., batch, flow-electrode, etc.). By controlling for the separation conditions, new materials,\cite{Huang2017} device geometries,\cite{Suss2015} electrode configurations,\cite{Mubita2018,Qu2018} and operation modes\cite{Qu2016,Dykstra2018,Garcia-Quismondo2016} can be objectively compared to one and other in terms of their energy consumption and volume throughput (see Section 2.1 and 2.2 for an example of such a comparison). 

Here we advocate for $E_v$ and $P$ as opposed to, for example, the various other metrics mentioned above because the end product of deionization is most often a volume of treated water, which means that volume-based performance measures are more directly relevant to practical applications. While it may seem that additional performance measures lend further insight, in fact many of these metrics are closely related by definition (see section 3.0.5). In the end, the total cost of the process is of course the ultimate comparator, but cost can vary too much due to a number of application-specific factors. Thus, here we use the productivity as a surrogate for capital cost, and the energy consumption as a surrogate for operational cost. In addition, the choice of $E_v$ and $P$ as primary performance measures will help facilitate comparisons between CDI and other desalination methods, better defining where and how CDI is technologically relevant.\cite{Pan2018,Zhao2013,VanLimpt2014} Finally, the approach of Fig. \ref{Fig: pictogram} has a direct relevance to thermodynamic efficiency, which is another insightful comparator of separation efficiency.\cite{Dugoecki2013,Wang2018a,Biesheuvel2009,Hemmatifar2018} Thus, from here we refer to $E_v$ and $P$ as primary performance metrics and the various other (molar) metrics as secondary performance metrics. 

In addition to $E_v$ and $P$ as primary performance measures, Fig. \ref{Fig: pictogram} also highlights the other crucial aspect of our perforamnce framework: defining the separation conditions. The issue with overlooking separation conditions is that virtually any desalination performance metric will be enhanced at lower water recovery and/or removal (i.e., lower concentration reduction or absolute amount of moles removed). Indeed, it has been demonstrated that SAC, ASAR, and ENAS can all improve when the magnitude of concentration reduction is decreased.\cite{Hawks2017,Wang2018} Fig. \ref{Fig: performance slice} also confirms this effect, showing that for higher $\left\langle \Delta c \right\rangle$ energy consumption increases and productivity decreases or remains constant. In general, improving any performance metrics without a set minimum concentration reduction will result in a maximum value at irrelevant, vanishingly small concentration reductions. Furthermore, the risk of only examining metrics normalized by salt adsorbed or concentration reduced is that such parameters mask the absolute amount removed. Concentration reduction is of course not the only important separation condition, as CDI performance is expected to increase with higher feed salinity due to lower ionic resistances. Thus, CDI performance results at the same concentration reduction and water recovery are not equivalent if taken with different feed concentrations. Finally, there is no straightforward way to convert performance metrics between separation conditions. Even for simple changes in the separation, a quantitative prediction of how performance should scale is difficult, requiring a full well-calibrated model.\cite{Ramachandran2018} If multiple parameters are changed, a quantitative comparison of experiment and data is even more difficult.\cite{Biesheuvel2014} 

\begin{figure}[h]
	\centering
	\includegraphics[width=0.8\columnwidth]{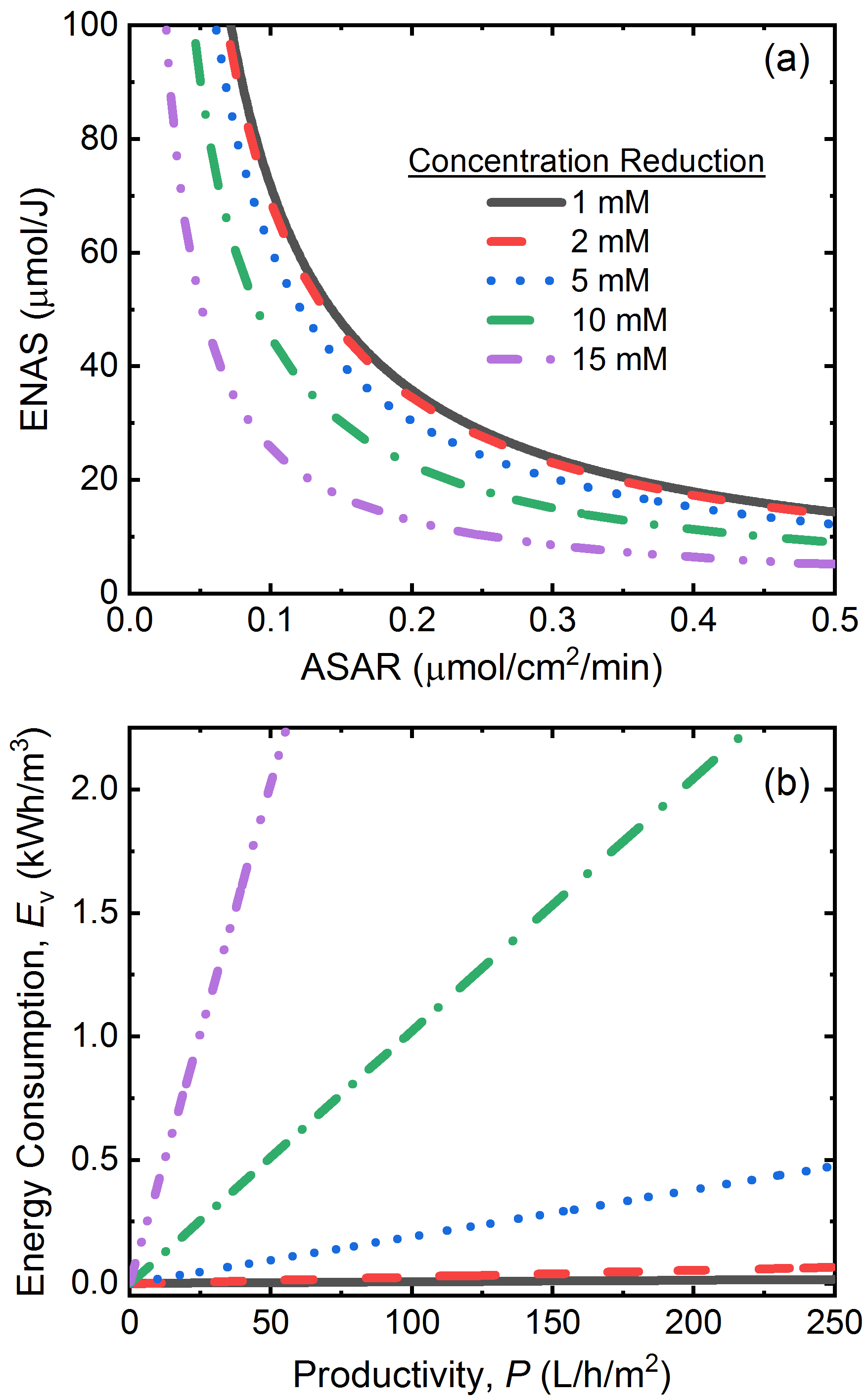}
	\caption{Example calculations of the CDI performance metrics (a) ENAS vs. ASAR and (b) \textit{E}$_v$ vs. \textit{P} for different average concentration reductions and \textit{WR} = 50\%, $c_{\text{\tiny feed}} = 50$ mM,  CC operation, a fixed cell voltage window corrected for ohmic drop, and using the model and parameters from Hawks et al.\cite{Hawks2017}. These simulations assume perfect energy recovery and reuse ($\eta = 1$) from the external circuit.}
	\label{Fig: ASAR vs ENAS}
\end{figure}

To illustrate the need for specifying separation conditions, in Fig. \ref{Fig: ASAR vs ENAS} we plot simulated performance curves from the simple resistive model of Hawks et al.\cite{Hawks2017} for various concentration reductions ($\left\langle \Delta c \right\rangle$). Fig. \ref{Fig: ASAR vs ENAS} shows that both removal rate (as described by ASAR), throughput (as described by productivity), and energy consumption (as described by ENAS or $E_v$) are simultaneously and monotonically enhanced at lower average concentration reduction of the feed stream. The input parameters for the model were taken from experimental results from Hawks et al.,\cite{Hawks2017} and the plot was generated by varying the applied current and flow rate for CC operation with a constant 1 V voltage window.
The model assumes no Faradaic losses, but this assumption will not change the overall conclusion that performance typically increases with lower concentration reduction. We note that the same general conclusions were recently reached by Wang et al.\cite{Wang2018} with a different cell geometry and device model. Thus, without specifying the concentration reduction, for example, one is free to exploit the fact that ``desalinated" water is any water that has a salt concentration less than the feed, regardless of how much less concentration. Additionally, both Fig. \ref{Fig: ASAR vs ENAS} and Fig. \ref{Fig: performance slice} highlight the inherent trade-off between throughput and energy consumption that often emerges, reinforcing the assertion that it is critical to report both energy and throughput metrics for a given separation.\cite{Wang2018} 
\FloatBarrier
\subsection{Example Standardized Separation for CDI Research}
\begin{figure*}[!h]
	\centering
	\includegraphics[width=1.7\columnwidth]{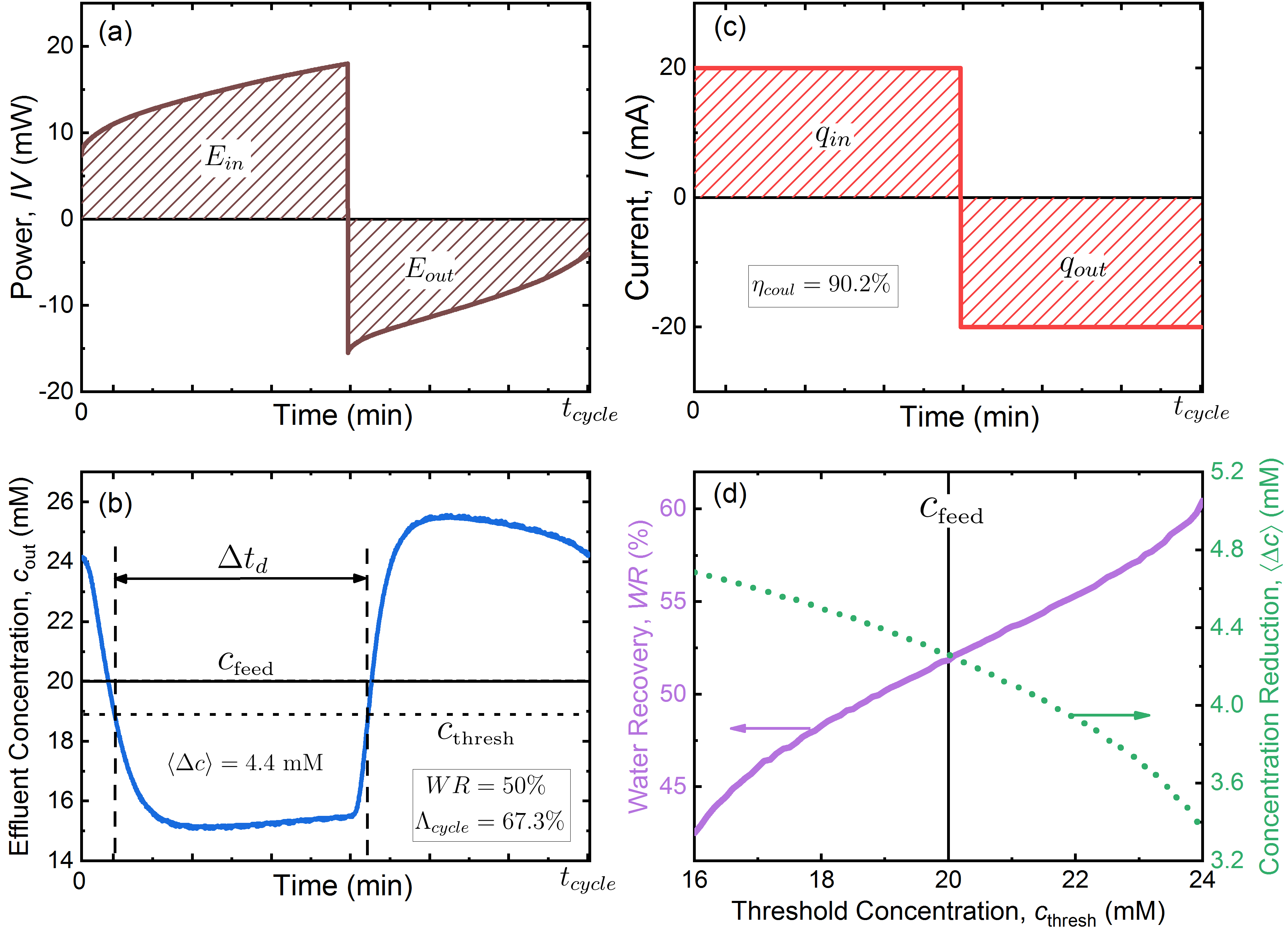}
	\caption{Experimental data curves for a fte-CDI cell under CC operation ($Q = 2$ ml/min, $A = 22.4$ cm$^2$) showing (a) power vs. time, (b) effluent salt concentration vs. time with a $c_{\text{\tiny thresh}} = 18.9$ mM, (c) current vs. time, and (d) the effect of threshold concentration on $WR$ and $\left\langle \Delta c \right\rangle$. The plot in (d) illustrates how a different choice of $c_{\text{\tiny thresh}}$ relative to the typical choice of $c_{\text{\tiny thresh}} = c_{\text{\tiny feed}}$ alters $\left\langle \Delta c \right\rangle$ and $WR$. This data underlies point in Fig. \ref{Fig: performance slice} marked by $\Theta$.}
	\label{Fig: Example analysis}
\end{figure*}  
\indent \indent For comparing CDI research results, we propose the following example separation condition as a nominal case: a feed water of 20 mM NaCl; an average concentration reduction $\left\langle \Delta c \right\rangle$ of 5 mM in dynamic steady-state (DSS) cycling operation;\cite{Porada2012,Zhao2013a,Hemmatifar2016} and a water recovery of 50\%. This means that half of the effluent water volume will have an averaged concentration of 15 mM, and the other half will have a concentration of 25 mM. Under these removal conditions, we propose reporting at least the two performance parameters mentioned above: volumetric energy consumption ($E_v$) and throughput productivity ($P$). 

For this separation, we propose a feed concentration of 20 mM NaCl because it is in the middle of the salinity range where CDI is expected to industrially relevant.\cite{Zhao2013,Pan2018,VanLimpt2014} We chose a $\left\langle \Delta c \right\rangle$ of 5 mM and a water recovery of 50\% because both of these are reasonably attainable in a research lab setting. It would be more relevant to examine a 15 mM removal from a 20 mM feed at a WR of 75\% or higher, but this requires significantly more engineering to achieve. Since a reporting standard will only be adopted if its reporting requirements are reasonably attainable, it would be counterproductive to propose removal conditions that are too difficult to meet. However, too small of a concentration reduction pertains more to salt adsorption performance than to desalination performance, which overestimates the suitability of the device for water treatment purposes. Thus, we propose a 5 mM removal here as a reasonable compromise between achievability and relevance to desalination.    

In order to obtain $E_v$ and $P$ at WR $= 50$\%, one can analyze the concentration vs. time data with respect to a concentration threshold ($c_{\text{\tiny thresh}}$), below which water is considered to be collected as diluate and above which water is considered to be collected as concentrate (Fig. \ref{Fig: Example analysis}b). In Fig. \ref{Fig: Example analysis} we show example data for constant flow rate constant current (CC) operation with a flow-through electrode (fte-CDI) device similar to that described in Ref. \citenum{Hawks2017} (see Table S2 in the supporting information (SI) for cell particulars). Not shown as example data but still analyzed for comparison is a flow between membrane CDI (fb-MCDI) device (see Table S3 in the SI for cell particulars). The typical choice of threshold concentration (if one is used) is the feed concentration ($c_{\text{\tiny thresh}} = c_{\text{\tiny feed}}$); however, this is arbitrary. Fig. \ref{Fig: Example analysis}b and \ref{Fig: Example analysis}d illustrate how choosing a different threshold concentration alters the time over which diluate is collected within a DSS cycle and subsequently how $\left\langle \Delta c \right\rangle$ and $WR$ trade off against each other. An example numerical approach for analyzing data like that in Fig. \ref{Fig: Example analysis} along with a calculation spreadsheet written in generally available commercial software (Microsoft Excel) are provided in the SI.
\begin{figure*}[!h]
	\centering
	\includegraphics[width=1.65\columnwidth]{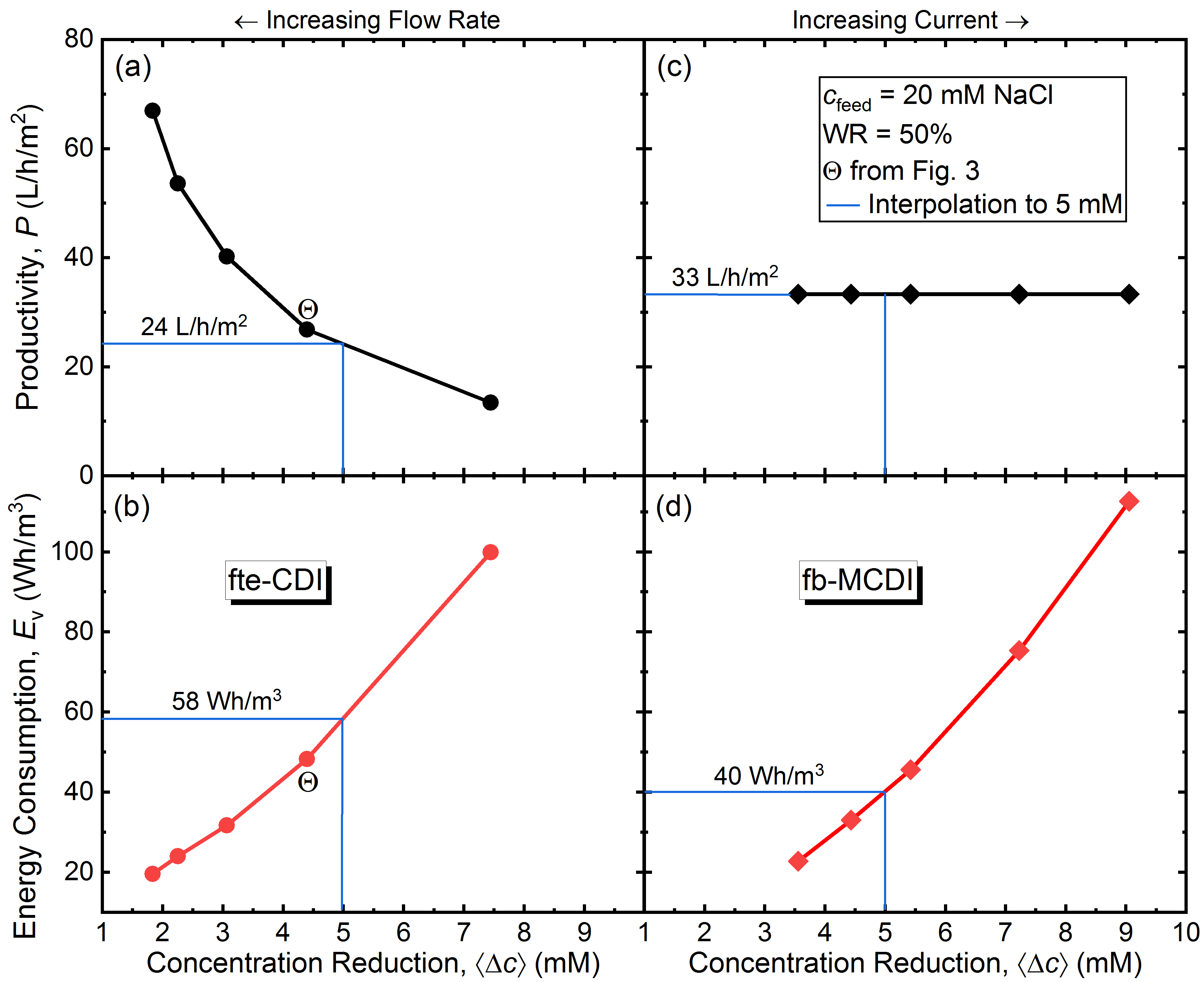}
	\caption{Example experimental performance traces of productivity (a,c) and energy consumption (b,d) for the fte-CDI cell (a,b) described in Fig. \ref{Fig: Example analysis} and for the fb-MCDI cell (c,d) described in the SI. The operational mode was DSS CC with a 20 mM NaCl feed solution and the \textit{WR} 50\% for both cells. Both energy calculations assume perfect energy recovery and reuse ($\eta = 1$) from the external circuit. The data underlying the point marked by $\Theta$ in a,b corresponds to Fig. \ref{Fig: Example analysis}. Raw data is presented in the SI.}
	\label{Fig: performance slice}
\end{figure*} 

In terms of measuring $E_v$ and $P$ for a given $\left\langle \Delta c \right\rangle$, one can extract values at the relevant concentration reduction by sweeping through an operational variable, plotting $P$ vs. $\left\langle \Delta c \right\rangle$ and $E_v$ vs. $\left\langle \Delta c \right\rangle$, and linearly interpolating at the $\left\langle \Delta c \right\rangle$ of interest (Fig. \ref{Fig: performance slice}). Curves of this type for the fte-CDI cell in Fig. \ref{Fig: performance slice}a,b were generated by varying the flow rate (1-5 ml/min) and holding a constant charging/discharging current (0.89 mA/cm$^2$) and voltage window (0.2-0.9 V). Similarly, for the fb-MCDI device in Fig. \ref{Fig: performance slice}c,d, such data was generated by varying the magnitude of the charging/discharging current (0.74-1.85 mA/cm$^2$) and holding a constant flow rate (3.75 ml/min per cell), discharge voltage (0 V), and charging time (3 min). The upper charging voltage was not controlled within the voltage window of 0.0-1.4 V. These example experimental sweeps give a range of performance and $\left\langle \Delta c \right\rangle$ values (see SI Fig. S3, Fig. S4, and Fig. \ref{Fig: Example analysis} for raw data). In this way, the 5 mM removal condition, for instance, can be readily estimated from a linear interpolation of the results. Note that we do not advocate for performance extrapolation, and interpolation should be conducted with reasonable data spacing. 

Notably, the energy consumption data in Fig. \ref{Fig: performance slice} assumes perfect capacitive energy recovery and reuse from the discharge, which is an idealized case. Due to the capacitive nature of CDI, a portion of the energy put into the desalination process is recoverable during the discharge; however, a real circuit that drives CDI devices will not transfer energy perfectly, but only recover a fraction ($\eta$) of the total recoverable energy added to the CDI cell during desalination. Importantly, CDI measurements in the lab virtually always measure the total recoverable energy, as opposed to actually use it to power another CDI device or the same device at a later time. Thus, in general, it is important to note energy consumption values assuming no energy recovery so to know the range of expected values if a real circuit was used. For Fig. \ref{Fig: performance slice}a,b, the 5 mM removal with no energy recovery ($\eta = 0$) consumes 150 Wh/m$^3$, or about 2.6$\times$ more energy per volume of diluate than with perfect energy recovery and reuse. 

As noted above, another benefit of using our proposed separation conditions and metrics is that energy consumption can be readily compared to thermodynamic calculations. For instance, calculating the specific Gibbs free energy of separation as defined in Wang et al.\cite{Wang2018a} for the 5 mM removal in Fig. \ref{Fig: performance slice}b reveals that the thermodynamic minimum $E_v$ is 2.8\% of the measured value of 58 Wh/m$^3$. In the same vein, the fb-MCDI cell in Fig. \ref{Fig: performance slice}d has a thermodynamic efficiency of 4.1\% as well as a larger productivity for the same separation (see SI Section 3 for full details of calculation and Fig. S6 and Fig. S7 for data).\cite{Dugoecki2013,Hemmatifar2018} This analysis gives context to the energy consumption values in terms of how much more efficient a separation could be if ideally carried out. Of course, even when such efficiencies are compared for equal separations (as in Fig. \ref{Fig: performance slice}), one must also weigh them against productivity to capture the full performance picture. 

In terms of device comparisons, Fig. \ref{Fig: performance slice} provides an example for how to objectively compare different desalination systems (e.g., here CC fte-CDI in (a,b) vs. CC fb-MCDI in (c,d)). As expected, for identical $\left\langle \Delta c \right\rangle = 5$ mM, $c_{\text{\tiny feed}} = 20$ mM, $WR = 50$ \%, $\eta = 1$ separations, the MCDI system shows significantly enhanced performance when compared to the membrane free fte-CDI system. Specifically, the fb-MCDI device consumes $\sim$69\% of the energy at 1.38$\times$ the productivity when compared to a fte-CDI cell. Importantly, such performance gains must be weighted against the additional cost that accompanies the use of membranes. In the next section we expand upon Fig. \ref{Fig: performance slice} by further varying operation conditions to achieve multiple performance points ($P$, $E_v$) at identical $\left\langle \Delta c \right\rangle = 5$ mM, $c_{\text{\tiny feed}} = 20$ mM, $WR = 50$ \%, $\eta = 0,1$ separations. In this way, a more comprehensive picture of performance space can be obtained.     
\FloatBarrier
\subsection{Performance Tradeoffs}
\begin{figure}[h]
	\centering
	\includegraphics[width=0.9\columnwidth]{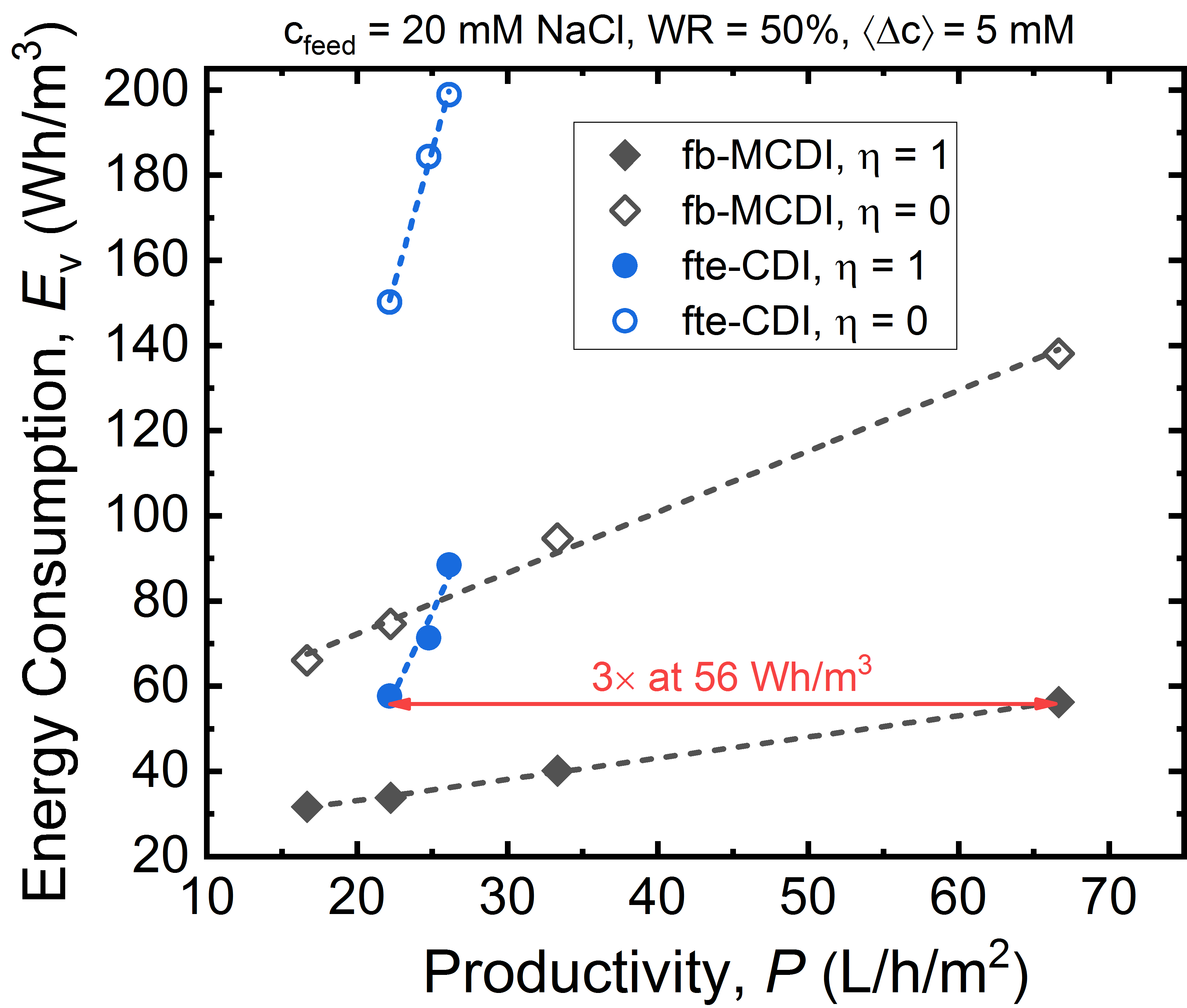}
	\caption{Performance tradeoff relationships for fb-MCDI and fte-CDI with perfect energy recovery and reuse and no energy recovery. The separation conditions are identical for every data point: $\left\langle \Delta c \right\rangle = 5$ mM, $c_{\text{\tiny feed}} = 20$ mM, $WR = 50$ \%, $\eta = 0,1$. These curves expand upon the results of Fig. \ref{Fig: performance slice} using the same parameters but further varying operational variables.}
	\label{Fig: Performance_Curves}
\end{figure}

\indent \indent For a more complete understanding of performance space, one can generate traces like those in Fig. \ref{Fig: performance slice} over several operational variables of interest (e.g., voltage window, charging time, charging current, flow rate, etc.) and then plot $E_v$ vs. $P$ for identical separations. In this vein, Fig. \ref{Fig: Performance_Curves} shows experimental results for the fte-CDI and fb-MCDI devices discussed above. Fig. \ref{Fig: Performance_Curves} further reveals the range of performance that can be expected when using control circuits that have perfect energy recovery and reuse ($\eta = 1$) and no energy recovery or reuse ($\eta = 0$). The CC operation data in Fig. \ref{Fig: Performance_Curves} for the fte-CDI device was generated by varying the flow rate and charging current and holding the voltage window constant, while the data for fb-MCDI device was obtained by varying flow rate and charging/discharging current and holding the charging time constant.

The relationship between $E_v$ and $P$ for the CC operation observed in Fig. \ref{Fig: Performance_Curves} agrees well with the model predictions of Fig. \ref{Fig: ASAR vs ENAS}b, all showing a clear linear tradeoff between throughput and energy consumption. The difference in performance between CC fte-CDI and CC fb-MDCI is further elucidated in Fig. \ref{Fig: Performance_Curves}, but again must be weighted against the additional cost that accompanies the use of membranes. The larger difference in energy consumption between $\eta = 0,1$ for the fte-CDI cell vs. the fb-MCDI cell implies that the fte-CDI system is storing a higher amount of the total energy put into the desalination cycle when compared to the fb-MCDI cell, likely due to the lower series resistance (see SI Tables S2 and S3).  
  
In terms of understanding how well an electrode material, device, or operation mode performs, one can compare productivity at equivalent energy consumption to assess the relative required system size at a similar operating cost. For example, in Fig. \ref{Fig: Performance_Curves} with $\eta = 1$, the fb-MCDI device has $\sim$3$\times$ more throughput than the fte-CDI cell at equivalent energy consumption ($\sim$56 Wh/m$^3$), implying that the fte-CDI cell must be at least roughly $\sim$3$\times$ less expensive than the fb-MCDI device to be competitive. Alternatively, one can compare energy consumption at equivalent productivity to get a sense of the difference in operating cost for a similar system size. Overall, these comparisons highlight how the performance framework introduced herein enables a detail-independent assessment of the CDI system's capabilities in terms of its economically relevant deionization performance.
\FloatBarrier
\section{Definitions}
\indent In this section, we mathematically define how to calculate the proposed performance metrics. Importantly, all analyses are intended for cyclic data taken over a full dynamic-steady state (DSS)\cite{Porada2012,Zhao2013a,Hemmatifar2016} charge/discharge cycle. 
To supplement these performance values, it is also important to publish the raw effluent concentration or conductivity vs. time data and the relevant current/voltage vs. time data for a DSS cycle (e.g., Fig. S3 in the supporting information). In our case, DSS is typically achieved on the third or fourth identical cycle of a given operation condition. Observing that the effluent conductivity begins and ends at the same value is a simple method for checking that DSS has been achieved. We also discuss below how these definitions must be modified for systems with flowable electrodes operating continuously rather than cyclically. Note that here and throughout the paper unit conversions are implied.
\subsubsection{Water Recovery}
\indent For the water recovery $WR$ (Fig. \ref{Fig: pictogram}), we define
\begin{align}
\label{Vdesal}
V_d &= \int_{\Delta t_d} Q~dt \\
\label{Vbrine}
V_c &= \int_{\Delta t_c} Q~dt \\
\label{WR}
WR  &= \frac{V_d}{V_d + V_c},
\end{align}
where $Q$ is the volumetric flow rate through the cell, $V_d$ is the volume of desalinated water (diluate) collected that, when mixed, has a lower salinity relative to the feed stream, $V_c$ is the volume of ``brine" water (concentrate) collected that, when mixed, is enriched in salt relative to the feed stream, $\Delta t_d$ is the time over which the desalinated water is collected, $\Delta t_c$ is the time over which the concentrate stream is collected, and $WR$ is the water recovery (\%). The typical choice of $\Delta t_d$ is everywhere where the effluent conductivity is less than the feed conductivity (Fig. \ref{Fig: Example analysis}b); however, the user is in fact free to choose the region $\Delta t_d$ over a DSS cycle arbitrarily so long as final average salinity is lower than the feed. In our definition, the total volume of fluid flowed over the DSS cycle is $V_d + V_c$.
\subsubsection{Concentration Reduction}
\indent The salt removed from the effluent stream and the average concentration reduction can be calculated from
\begin{align}
\label{Delta Salt}
\Delta N_{d} &= \int_{\Delta t_d} Q (c_{\text{\tiny feed}} - c_{out}) dt  \\
\Delta N_{c} &= \int_{\Delta t_c} Q (c_{\text{\tiny feed}} - c_{out}) dt  \\
\label{deltaC bar}
\left\langle \Delta c \right\rangle  &= \frac{\Delta N_{d}}{V_d},
\end{align}
where $\Delta N_{d}$ is the salt removed over one cycle relative to the feed stream as measured from the effluent (moles), $\Delta N_{c}$ is the salt added to the concentrate stream over one cycle as measured by the effluent (moles),$c_{out}$ is the effluent salt concentration (mM), and $\left\langle \Delta c \right\rangle$ is the average concentration reduction (mM). $\Delta N_{d}$ is maximum for the choice of $\Delta t_d$ such that $c_{\text{\tiny thresh}} = c_{\text{\tiny feed}}$, and will go down if a smaller portion of more strongly desalinated water is kept to enhance concentration reduction, or if a portion of the concentrate is used to enhance water recovery (Fig. \ref{Fig: Example analysis}d). 
\FloatBarrier
\subsubsection{Energy Consumption}
\indent For volumetric energy consumption per cell ($E_v$) of Fig. \ref{Fig: pictogram}, we propose
\begin{align}
E_{in} &= \int_{\Delta t_{cycle}} IVdt~~\text{where ~} IV > 0 \\
E_{out} &= \int_{\Delta t_{cycle}} IVdt~~\text{where ~} IV < 0 \\
\label{Wh/m3}
E_v &= \frac{E_{in} - \eta E_{out}}{V_d} \\
\label{kJ/mol}
E_m &= \frac{E_{in} - \eta E_{out}}{\Delta N_{d}}
\end{align}
where $IV$ is the current-voltage product for a single electrode pair (W), $\Delta t_{cycle}$ is the total DSS cycle time (min), $E_{in}$ is the total energy input during the DSS cycle (J), $E_{out}$ is the total recoverable energy from the cell over the DSS cycle (J), $E_v$ is the volumetric energy consumption (Wh/m$^3$), $E_m$ the molar energy consumption (kJ/mol), and $\eta$ is the fraction of $E_{out}$ actually recovered and reused to power another charging phase. Thus, if perfect energy recovery is assumed as in Fig. \ref{Fig: performance slice}, then implicitly $\eta = 1$. Examples of practical values of $\eta$ can be found in the literature.\cite{Landon2013a,Kang2016,Pernia2014,Pernia2012} Since real-world energy transfer circuits (e.g., buck-boost converters) are not perfectly efficient, it is important to report $E_v$ for at least two values of $\eta$ (e.g., 0 and 1). If there are multiple devices in series or parallel, then $E_{in}$, $\eta E_{out}$, and ${V_d}$ in the Eqn. \eqref{Wh/m3} must be summed appropriately. A graphical example of $E_{in}$, $E_{out}$, and $\Delta t_{cycle}$ for the cell examined herein is shown in Fig. \ref{Fig: Example analysis}a. If the pumping energy is significant compared to the desalination energy $E_v$, then this number should be reported as well. The pumping energy for a CDI cycle can be calculated as $E_p = \int pQ dt/V_d$ (Wh/m$^3$) where $p$ is the pressure drop across the cell (psi) and the integral is over a cycle. For the data in Fig. \ref{Fig: performance slice}a and \ref{Fig: performance slice}b, the pumping energy is at worst 4$\times$ lower than the desalination energy ($\sim$5 Wh/m$^3$), making it safe to assume that it is negligible.  

\subsubsection{Productivity}
\indent For throughput we use the term productivity (Fig. \ref{Fig: pictogram}),\cite{Pan2017} which is given by
\begin{align}
\label{Producitivity}
P &= \frac{V_d}{n \cdot A \cdot  \Delta t_{cycle}}
\end{align}
where $P$ is the productivity in L/h/m$^2$, $n$ is the number of cells (electrode pairs) in the CDI stack (series and parallel), and $A$ is the projected face area (cm$^2$) of one cell as discussed above and illustrated in Fig. S2 in the SI. The productivity is simply a volume of diluate produced per unit time per projected face area of device. Since the separation condition (i.e., $\left\langle \Delta c \right\rangle $, $WR$, and $c_{\text{\tiny feed}}$) is already specified, there is only a need to specify volume throughput and not an average removal rate like with ASAR. Normalizing by projected face area rather than electrode mass is more directly related to cell resistance and cost when membranes are used. Similarly, volume throughput for a given separation is more closely tied to general desalination performance as opposed to salt removal rate,\cite{Ghaffour2013,Zhao2013} which is more specific to CDI-type desalination techniques. 

\subsubsection{Performance Metric Relationships}
\indent Finally, it is important to understand the defined relationships between traditional performance metrics---what we call here secondary performance metrics---and those detailed above. For instance, recalling that the previous performance metrics ASAR and ENAS as defined in Ref. \citenum{Hawks2017} are 
\begin{align}
\label{ASAR}
ASAR &= \frac{\Delta N_{d}}{n \cdot A \cdot  \Delta t_{cycle}} \\
\label{ENAS}
ENAS &= \frac{\Delta N_{d}}{E_{in} - \eta E_{out}}, 
\end{align}
then comparing Eqns. \eqref{ASAR} and \eqref{ENAS} to Eqns. \eqref{deltaC bar} and (\ref{Wh/m3}-\ref{Producitivity}) reveals that 
\begin{align}
\label{P and ASAR}
P &= \frac{ASAR}{\left\langle \Delta c \right\rangle} \\
\label{Ev and ENAS}
E_v &= \frac{\left\langle \Delta c \right\rangle}{ENAS} \\
E_m &= \frac{E_v}{\left\langle \Delta c \right\rangle}. 
\end{align}
Furthermore, combining Eqns. \eqref{P and ASAR} and \eqref{Ev and ENAS} gives
\begin{align}
P \cdot E_v &= \frac{ASAR}{ENAS}.
\end{align}
Thus, with $\left\langle \Delta c \right\rangle$ specified, the volumetric metrics proposed here can be converted to the traditional molar parameters and vice versa. Note that here and throughout the paper unit conversions are implied. 

The exercise of examining how various performance metrics in the CDI literature are related by definition is important, because it reveals where metric comparisons do not necessarily give additional insight (only relations by definition). Ultimately, we chose to advocate for the two parameters $E_v$ and $P$ because they reflect  generally desirable desalination characteristics for any system, regardless of its operating mechanism. We do not advocate using molar-based metrics because the end product of deionization is a volume of treated water, not a mole of salt, which makes volume-normalized quantities of a more broad interest. Interestingly, though, ENAS by definition is $ENAS = \left\langle \Delta c \right\rangle/E_v$, which gives a sense of how efficiently a given $\left\langle \Delta c \right\rangle$ is achieved; however, this number must be analyzed extremely carefully because the best values can often occur in a region of separation space that is irrelevant to practical applications (i.e., low $\left\langle \Delta c \right\rangle$). Thus, various normalized analysis of desalination performance that deviates from the Fig. \ref{Fig: pictogram} framework must be conducted with caution. 
 
\subsubsection{Non-DSS Operation}
While DSS cycling is the dominant mode of operation for CDI, there are other operation modes like batch-mode and flow-electrode setups that require a different approach to calculating the various parameters described above. In general, though, the approach depicted in Fig. \ref{Fig: pictogram} holds regardless of operation method.   

For systems with flowable electrodes,\cite{Doornbusch2016} the definitions Eqns. \eqref{Vdesal}-\eqref{Producitivity} still hold by replacing the specified integration time with the total experiment time. For batch mode operation,\cite{Porada2013} the individual quantities and $\left\langle \Delta c \right\rangle$, ${V_d}$, $E_{in}$, $\eta E_{out}$ along with system size and experiment time can be used to directly compute the values of Fig. \ref{Fig: pictogram}. In essence, the equations above for DSS operation can be adapted straightforwardly to give values that faithfully reflect those of Fig. \ref{Fig: pictogram}.  

\section{Performance Indicators}
\indent In this section, we will discuss parameters which we believe are highly useful for understanding cell performance and material limitations, but are not performance metrics in and of themselves. These parameters combined with the reported operational performance metrics above can help better understand why certain materials, device architectures, and operation modes are more suitable for CDI, or in contrast, are less suitable. If the primary goal of the study is to evaluate a novel electrode material, then such extra characterization is especially critical. Importantly, the following descriptions are not meant to be comprehensive technical reviews, but rather are included to lend context to the CDI performance metric/indicator landscape.  

\subsection{Salt Adsorption Capacity}
\indent We begin our performance indicator discussion with the salt adsorption capacity (mg/g), which for a CDI cell over a DSS cycle is given by
\begin{align}
\label{SAC}
SAC &= \frac{M}{m} \Delta N_{d} 
\end{align}
where $M$ is the molar mass of the salt (g/mol), $m$ is the total mass of electrodes in the cell (g). The SAC is a function of operational flow efficiency (defined below and in Ref. \citenum{Hawks2017}) and magnitude of adsorbed salt over a DSS cycle.\cite{Hawks2017} This is in contrast to the maximum salt adsorption capacity (mSAC) or identically equilibrium salt adsorption capacity (Eq-SAC), which are taken at constant-voltage equilibrium. The Eq-SAC value is best measured over many charging voltages to at least $V = 1.2$ V (with the discharge voltage typically set to zero), a salinity of 20 mM, and using NaCl so that experiments can be compared between different studies.\cite{Guyes2017,Suss2015}. Notably, if the electrode material density is reported ($\rho_e$), then different Eq-SAC normalizations (e.g. mg/ml or mol/ml) can be readily evaluated.

The strength of Eq-SAC is that it is an indicator of the upper limit of desalination capacity over a given voltage window, where higher values typically correspond to larger obtainable removals during operation (i.e., higher possible concentration reductions and water recoveries). However, Eq-SAC is not a desalination performance metric in and of itself, and therefore cannot be used as a substitute for energy consumption and throughput. Eq-SAC is influenced by a combination parameters, including: equilibrium charge efficiency, charging and discharging voltages, point of zero charge (PZC), and material capacitance.  

\subsection{Capacitance and PZC}
\indent \indent The capacitance and PZC are important factors that underly SAC, and therefore also strongly impact the desalination capacity of a cell.  A discussion of capacitance characterization for electrochemical supercapacitors is given by Zuliani et al.\cite{Zuliani2015}, and analysis of the material point of zero charge (PZC) and its influence on CDI performance can be found in Refs. \citenum{Avraham2011a,Gao2015,Gao2018,Omosebi2015,Omosebi2014,Hemmatifar2017}. 

In brief, though, the cell's differential capacitance can be evaluated through the use of the equation
\begin{align}
\label{Capacitance}
C &= I \left( \frac{dV}{dt} \right) ^{-1}
\end{align}
where $V$ is the cell voltage (V), $I$ is the current (A), and $C$ is the capacitance (F), which can be converted to material capacitance (F/g or F/cm$^3$) by considering the electrodes in a CDI cell as two capacitors in series.\cite{Zuliani2015} In terms of normalization,\cite{Zuliani2015,Gogotsi2011} the electrode volumetric capacitance in F/cm$^3$ is more relevant to CDI operation,\cite{Hawks2017} but the gravimetric capacitance in F/g has better comparability to electrochemical capacitors at large; thus, assessing both is important. Similar to SAC, if the electrode material density is reported, then the volumetric and gravimetric capacitance values can be calculated from one another.  Eqn. \eqref{Capacitance} suffices for capacitance assessment of both supercapacitors and pseudocapacitors;\cite{Brousse2015} however, it may not be the best way to characterize intercalation storage mechanisms, which are an emerging area of interest in CDI.\cite{Porada2017,Kim2017a,Suss2018} In the case of charge storage by intercalation, electrode capacity in mAh/cm$^3$ and mAh/g and voltage window are perhaps more appropriate.

In contrast to energy-storage studies, the relevant material capacitance can be difficult to evaluate due to the low electrolyte concentrations pertinent to CDI operation (e.g., 20 mM NaCl). At these low concentrations, significant PZC, depletion, resistive, and Faradaic effects can obscure the capacitance evaluation by Eqn. \eqref{Capacitance}. We find that in CC operation, for example, the discharge portion of the desalination cycle can often provide a somewhat reasonable estimation of $C$ through Eqn. \eqref{Capacitance}. This analysis is identical to evaluating the capacitance by the method of galvanostatic cycling, which is a common approach for capacitance evaluation.\cite{Zuliani2015} Another way to measure capacitance is with cyclic voltammetry (CV) taken at concentrations and (multiple) sweep rates $dV/dt$ that are comparable to the ones used for desalination. For such a CV measurement, the electrode material PZC can be estimated from the voltage at which the capacitance from Eqn. \eqref{Capacitance} passes through a local minimum. For purely comparative purposes, the differential capacitance can be taken as the value of $C$ at $V = 0$ ($C_{V=0}$, see SI Fig. S9). An illustration of these effects and CV analysis are presented in the SI Fig. S9, including for the cell from Figs. \ref{Fig: Example analysis} and \ref{Fig: performance slice} (Fig. S1). If the CV is measured in a CDI cell, then it can be done with sufficient flow rate such that depletion effects are avoided. It is important that the CV is measured both over enough cycles such that DSS is reached and at relevant concentrations since double-layer capacitance scales non-linearly with electrolyte concentration in the low ionic strength regime that is relevant to CDI. 

Other ways to measure the capacitance are by the galvanometric intermittent titration technique (GITT) or amperometric intermittent titration technique (AITT), which involve transferring a precise quantity of charge and measuring the change in voltage, or changing the cell voltage and measuring the charge transferred, respectively. These methods have been routinely applied in CDI for carbon-based materials,\cite{Zhao2010a,Porada2012,Porada2013,Porada2013a} as well as recently applied for CDI with intercalation materials.\cite{Porada2017} Compared to CV, GITT may be more applicable to intercalation materials,\cite{Porada2017} and AITT might be more practical at low salinities (e.g., 5 mM NaCl).\cite{Porada2012,Guyes2017,Kim2015}   

\subsection{Charge and Coulombic Efficiencies}
\indent \indent A further set of relevant performance indicators are the charge efficiency and Coulombic efficiency, which have been the subject of much research in CDI.\cite{Shanbhag2016,Zhao2012,Zhao2010a,Avraham2009,Avraham2010,Gao2014,Kim2015,Hemmatifar2016} In terms of DSS cycling, the charge transferred, total cycle charge efficiency $\Lambda_{cycle}$, and round-trip Coulombic efficiency $\eta_{coul}$\cite{Smith2014,Shanbhag2016} are defined here by 
\begin{align}
\label{qin}
q_{in} &= \int_{\Delta t_{cycle}} I dt~~\text{where ~} I > 0 \\
\label{qout}
q_{out} &= \int_{\Delta t_{cycle}} I dt~~\text{where ~} I < 0 \\
\label{Charge Eff Meas}
\Lambda_{cycle} &= \frac{F \Delta N_{d}}{q_{in}} \\
\label{RT Coulombic}
\eta_{coul} &= \frac{q_{out}}{q_{in}}
\end{align}
where $q_{in}$ is the charge transferred at positive current (C), $q_{out}$ is the charge transferred at negative current (C), and $F$ is Faraday's constant (96485 C/mol). Of the two periods of opposing current direction in Eqns. (19) and (20), we define positive current to be the ``charging step", which is that period where the energy input was higher than in the period of opposing current sign. Alternatively, if the voltage polarity does not change and one assumes charge conservation over a DSS cycle, then the charge transferred to the cell must not be greater than the charge released from the cell, or $q_{out} \leq q_{in}$. From this, the definition of positive current in Eqn. (19) also follows. A graphical example of $q_{in}$ and $q_{out}$ for the cell examined herein is shown in Fig. \ref{Fig: Example analysis}c. 

It should also be noted that there is some ambiguity in the literature regarding the use of $q_{in}$ or $q_{out}$ in Eqn. \eqref{Charge Eff Meas}. For instance, $\Lambda_{cycle}$ as defined here was referred to as the ``dynamic charge efficiency" in Refs. \citenum{Hawks2017} and \citenum{Zhao2012}, but in Ref. \citenum{Zhao2010a} the ``charge efficiency" was taken as $\Lambda_{cycle}/\eta_{coul} = F \Delta N_{d}/q_{out}$. The replacement of $q_{in}$ with $q_{out}$ in Eqn. \eqref{Charge Eff Meas} can be done to partly factor out Coulombic losses in the total cycle charge efficiency. While the true capacitive charge transferred over a cycle lies somewhere between $q_{in}$ and $q_{out}$, using the charge released is a reasonable approach as the capacitive charge transferred is generally expected to be closer to $q_{out}$. 

In general, these efficiencies reflect the disparity between transferred electrical charge on one hand and salt deficiency or excess in the effluent stream on the other. In typical unfunctionalized carbon-based electrodes, there are broadly three main reasons behind such disparities. First, at low potentials across the electric double layers, co-ion expulsion occurs to a significant extent compared to counterion adsorption, leading to inefficient net adsorption.  Second, unwanted parasitic reactions can consume accumulated electric charge via Faradaic reactions at one or more electrodes. And finally, a portion of the desalinated water inside the cell is often not recoverable due to operational effects.\cite{Hawks2017,Shang2017} Such effects are captured by the flow efficiency parameter,\cite{Hawks2017,Johnson1971} which is defined as the ratio of moles of salt removed from the effluent stream to the moles of salt adsorbed by the electrodes over a DSS cycle.\cite{Hawks2017} Overall, the total cycle charge efficiency is influenced by flow efficiency, double layer, and Faradaic effects,\cite{Hawks2017} and the round-trip Coulombic efficiency $\eta_{coul}$ of Eqn. \eqref{RT Coulombic} measures charge loss due to Faradaic processes.\cite{Bazant2013,Pillay1996,Shanbhag2016} Taken together, a full analysis of Eqns. \eqref{Charge Eff Meas} and \eqref{RT Coulombic} can yield significant insight as to where operational performance losses are being incurred.  

\subsection{Series Resistance}
\indent \indent Finally, since CDI cells are high-current low-voltage devices, the series resistance is an important parameter that can help understand performance.\cite{Qu2015,Qu2016,Dykstra2016} Energy loss in CDI occurs via primarily two processes: energy dissipation through Joule heating by current through the resistive elements in the system, and unwanted Faradaic reactions. The resistive elements can be categorized as a series resistance and an ionic resistance within the electrode pores. This series resistance includes external leads, contact resistances at current collectors, collector resistance (often negligible), and ionic resistance within the separator. The impedance across the electrode thickness can be thought of as a distributed series/parallel arrangement of ionic resistances, capacitive surfaces, and electronic resistance of the electrode material (i.e., a transmission line).\cite{Posey1966,Dunn2000,DeLevie1963} While the series resistance clearly affects energy consumption,\cite{Qu2016,Hemmatifar2016} it also influences operational performance through, for example, the $RC$ charging time in CV operation and voltage window in CC operation. 

Here we define the total series resistance ($R_s$) as the external electronic resistance $R_{\text{\tiny \textit{EER}}}$ plus the separator resistance $R_{sp}$ (i.e., $R_s = R_{\text{\tiny \textit{EER}}} + R_{sp}$).\cite{Dykstra2016} In other words, the series resistance is the resistance of every part of the system except the resistance in the porous electrodes. The external electronic resistance $R_{\text{\tiny \textit{EER}}}$ includes the resistance due to the current collector, wires, and contacts,\cite{Qu2015,Dykstra2016} whereas the separator resistance $R_{sp}$ captures the ionic resistance across the separator according to 
\begin{align}
\label{Rsp}
R_{sp} &= \frac{l_{sp} \tau_{sp}}{A \kappa p_{sp}},
\end{align}
where $l_{sp}$ is the separator thickness ($\mu$m), $\tau_{sp}$ is the separator tortuosity, $\kappa$ is the electrolyte conductivity in the separator (mS/cm), and $p_{sp}$ is the separator porosity. There is also a significant ionic transport resistance associated with the electrolyte in the macropores; however, this resistance is strongly frequency dependent, being an electrical short at high frequency that ideally transitions to a capacitance in series with the electrolyte resistance at low frequency.\cite{Suss2013,Posey1966,Dunn2000,DeLevie1963} 
\begin{table*}[h]
	\caption{Parameter Summary and Classification.}
	\label{tbl:summary}
	\raggedleft
	\begin{tabular}{|c|c|c|m{7cm}|}
		\hline
		Category & Symbol & Typical Unit & Definition \\
		\hline
		\multirow{3}{*}{\makecell{\\ \\ Separation Condition}}
		& $\left\langle \Delta c \right\rangle$ & mM & Average concentration reduction of the feed over a cycle. \\
		\cline{2-4}
		& $c_{\text{\tiny feed}}$ & mM & Salt concentration of the feed stream. \\
		\cline{2-4}
		& $WR$ & \% & The ratio of diluate volume to total volume of water flowed over a cycle. \\
		\hline
		\multirow{3}{*}{\makecell{Primary \\ Performance Metric}}
		& $E_v$ & Wh/m$^3$ or Wh/m$^3$ & Volumetric energy consumption reported for multiple energy recoveries $\eta$. \\
		\cline{2-4}
		& $P$ & L/h/m$^2$ & The volume of diluate per total cycle time per cell face area. \\
		\hline
		\multirow{3}{*}{\makecell{\\ Secondary \\ Performance Metric}}
		& $E_m$ & kJ/mol & Molar energy consumption. \\
		\cline{2-4}
		& $ENAS$ & $\mu$mol/J & Energy normalized adsorbed salt. \\
		\cline{2-4}
		& $ASAR$ & $\mu$mol/min/cm$^2$ & Average salt adsorption rate per cell face area. \\
		\hline
		\multirow{3}{*}{\makecell{\\ \\ \\ Performance Indicator}}
		& $C$ & F, F/cm$^3$, F/g & Cell and material capacitance with measurement method and salinity noted. \\
		\cline{2-4}
		& $\Lambda_{cycle}$ & \% & Total cycle charge efficiency. \\
		\cline{2-4}
		& $\eta_{coul}$ & \% & Coulombic efficiency of a cycle. \\
		\cline{2-4}
		& $R_s$ & $\Omega \cdot $cm$^2$ & Series resistance, the sum of the external electronic resistance $R_{\text{\tiny \textit{EER}}}$ and the separator resistance $R_{sp}$. \\
		\hline
		\multirow{3}{*}{\makecell{\\ \\ \\ \\ Cell Characteristic}}
		& $n$ & -- & Number of cells used in the separation. \\
		\cline{2-4}
		& $n_m$ & -- & Number of membranes used per cell. \\
		\cline{2-4}
		& $A$ & cm$^2$ & Projected face area of a CDI cell (i.e., face area over which electrodes overlap). \\
		\cline{2-4}
		& $l_e$ & $\mu$m & Electrode thickness. \\
		\cline{2-4}
		& $l_{sp}$ & $\mu$m & Separator thickness. \\
		\cline{2-4}
		& $m$ & g & Total mass of electrodes. \\
		\cline{2-4}
		& $\rho_e$ & g/cm$^3$ & Electrode mass density. \\
		\hline
	\end{tabular}
\end{table*} 

In terms of experiments, the series resistance in particular can be readily measured with electrochemical impedance spectroscopy (EIS) at high frequency (e.g., 10-100 kHz) by taking the value of the real part of the impedance as the imaginary part tends to zero (see SI Fig. S10 and S11). It is important to note that high frequency EIS measurements are capable of measuring the true $R_s$, whereas methods based on applying an abrupt current or voltage step and measuring the instantaneous response often captures a significant amount of electrode ionic resistance (unless high time-resolution equipment is used). Because Eqn. \eqref{Rsp} is dependent on ion concentration, $R_s$ will change dynamically throughout a CDI cycle due to changes in electrolyte conductivity from desalination and regeneration.\cite{Hemmatifar2016} Nevertheless, EIS measurements are useful for establishing a baseline for comparing different device architectures and better understanding differences in performance. Examples and illustrations of how to measure $R_s$ with EIS are presented in Section 6 of the SI. 

\section{Parameter Summary}
\indent \indent A summary of the essential separation conditions, performance metrics, and performance indicators that are discussed above are presented in Table \ref{tbl:summary}. In the SI, a table is given defining all parameters used in this work, including those in Table \ref{tbl:summary}. We also add to Table S2 and S3 several essential cell characteristics that are highly valuable in interpreting desalination results. For example, the dry electrode material density is a critical electrode property that allows for conversion between

gravimetric and volumetric parameters. Additionally, the number of membranes used per cell is important for estimating the overall device cost, as the membranes are typically a cost-driving factor for the devices that use them. Finally, general device geometric parameters like the electrode and separator thickness (dry, uncompressed) are important basic cell characteristics that influence the magnitude of many of the discussed parameters. The parameters in Table \ref{tbl:summary} for the cell that underlies the results in Fig. \ref{Fig: performance slice} along with other important cell characteristics are summarized in a separation report in the SI Table S2. For convenience, we also provide this report format in the form of a calculation spreadsheet written in generally available commercial software (Microsoft Excel) that can be downloaded and readily used for other studies.

\section{Conclusions}
\indent In conclusion, we proposed a framework for quantifying CDI performance in terms of volumetric energy consumption (Wh/m$^3$) and throughput productivity (L/h/m$^2$). We demonstrated that reporting these metrics for equivalent separation conditions is essential for comparability and obtaining practically meaningful values. For research purposes, we proposed a nominal standard removal of 5 mM out of a 20 mM NaCl feed stream at 50\% water recovery for comparing new materials, devices, and operation modes. Using this separation, we compare performance between a fte-CDI cell and fb-MCDI cell, showing how significantly different systems can be compared in terms of generic desalination performance. The need for introducing such a framework arises from previously used metrics being reported without specifying the removal conditions (concentration reduction, feed concentration, and water recovery). In general, metric normalization must be carefully considered so to not allow evaluation at unspecified removal. The rationale presented here resolves this issue and better aligns CDI performance reporting with the field of desalination in general. Finally, for context, we discuss various cell characteristics and performance indicators that are highly useful for understanding why certain performance metrics were measured. Taken together, this work helps drive CDI in a direction towards more comparable and relevant performance evaluations so that this technology can realize its full potential.   
\FloatBarrier
\begin{acknowledgement}
Work at LLNL was performed under the auspices of the US DOE by LLNL under Contract DE-AC52-07NA27344. S.P. acknowledges financial support by the Dutch Technology Foundation STW, which is part of the Netherlands Organization for Scientific Research(NWO), and which is partly funded by the Ministry of Economic Affairs (VENI grant no. 15071). All other work was supported by LLNL LDRD 18-ERD-024.
\end{acknowledgement}

\bibliography{Library}


\section*{Supplementary material}

# Supporting Information for: Performance Metrics for the Objective Assessment of Capacitive Deionization Systems

Steven A. Hawks,[†] Ashwin Ramachandran,[‡] Slawomir Porada,[¶,⊥] Patrick G. Campbell,[†] Matthew E. Suss,[§] P.M. Biesheuvel,[¶] Juan G. Santiago,[∥] and Michael Stadermann[∗,†]

[†]*Lawrence Livermore National Laboratory, 7000 East Avenue, Livermore, CA 94550, United States*
[‡]*Department of Aeronautics & Astronautics, Stanford University, Stanford, CA 94305, United States*
[¶]*Wetsus, European Centre of Excellence for Sustainable Water Technology, Oostergoweg 9, 8911 MA, Leeuwarden, The Netherlands*
[§]*Faculty of Mechanical Engineering, Technion-Israel Institute of Technology, Haifa, Israel*
[∥]*Department of Mechanical Engineering, Stanford University, Stanford, CA 94305, United States*
[⊥]*Soft Matter, Fluidics and Interfaces Group, Faculty of Science and Technology, University of Twente, Meander ME 314, 7500 AE Enschede, The Netherlands*

E-mail: stadermann2@llnl.gov

# 1 Parameter Definitions

Below is a table of the parameters used in the main text and SI along with their corresponding definitions and typical units.

Table S1: Parameter Definitions

| Parameter Symbol | Definition | Typical Unit |
|---|---|---|
| $c_{\text{feed}}$ | inlet/feed/influent concentration | mM |
| $c_{out}$ | outlet/effluent concentration | mM |
| $c_{\text{thresh}}$ | threshold concentration below which water is kept as diluate | mM |
| $\langle \Delta c \rangle$ | average concentration reduction of the desalinated effluent over a DSS cycle relative to $c_{\text{feed}}$ | mM |
| $V_d$ | volume of diluate produced in a DSS cycle | ml |
| $V_c$ | volume of concentrate produced in a DSS cycle | ml |
| $WR$ | water recovery | % |
| $E_v$ | volumetric energy consumption | Wh/m$^3$ |
| $E_{v,th}$ | thermodynamic minimum volumetric energy consumption for a separation | Wh/m$^3$ |
| $E_m$ | molar energy consumption | kJ/mol |
| $E_p$ | pump energy consumption | Wh/m$^3$ |
| $P$ | throughput productivity | L/h/m$^2$ |
| $\Delta N_d$ | measured salt removed over a DSS CDI cycle from effluent | moles |
| $\Delta N_c$ | measured salt enrichment over a DSS CDI cycle from effluent | moles |
| $Q$ | volumetric flow rate | ml/min |
| $\Delta t_d$ | time over which the desalinated water is collected | min |
| $\Delta t_c$ | time over which the concentrate stream is collected | min |
| $\Delta t_{cycle}$ | cycle time, $t_{cycle} = \Delta t_d + \Delta t_c$ | min |
| $F$ | Faraday's Constant = 96485 | C/mol |
| $A$ | projected face area per cell | cm$^2$ |
| $C$ | differential capacitance of the CDI cell | F |
| $R_{sp}$ | ionic resistance of the separator | $\Omega$-cm$^2$ |
| $R_{EER}$ | external electronic resistance, including the resistance due to the current collector, wires, and contacts | $\Omega$-cm$^2$ |
| $R_s$ | series resistance, $R_s = R_{EER} + R_{sp}$ | $\Omega$-cm$^2$ |
| $E_{in}$ | total energy expended in a DSS cycle | J |
| $E_{out}$ | total recoverable energy from a DSS cycle | J |
| $\eta$ | the fraction of $E_{out}$ recovered and used to power another charging step | % |
| $ASAR$ | average salt adsorption rate | $\mu$mol/(min·cm$^2$) |
| $ENAS$ | energy-normalized adsorbed salt | $\mu$mol/J |
| $SAC$ | cycle salt adsorption capacity | mg/g |

| Parameter Symbol | Definition | Typical Unit |
|---|---|---|
| Eq-SAC | maximum salt adsorption capacity of the cell | mg/g |
| $M$ | molar mass of the feed stream salt | g/mol |
| $m$ | mass of all electrodes | g |
| $n$ | number of electrode pairs in the CDI stack used for separation | – |
| $n_m$ | total number of membranes used per cell | – |
| $q_{in}$ | total charge transferred during charging in DSS cycling | C |
| $q_{out}$ | total charge transferred during discharging in DSS cycling | C |
| $\Lambda_{cycle}$ | total cycle charge efficiency | % |
| $\eta_{coul}$ | measured round-trip cycle Coulombic efficiency | % |
| $I$ | applied current during charging/discharging | mA |
| $V$ | applied voltage | V |
| $p$ | pressure drop across a cell | psi |
| $l_e$ | electrode thickness | µm |
| $l_{sp}$ | separator thickness | µm |
| $\rho_e$ | electrode mass density | g/cm$^3$ |
| $\sigma_e$ | electrode material electrical conductivity | S/cm |
| $p_{mA}$ | electrode macroporosity | % |
| $p_{mi}$ | electrode microporosity | % |
| $p_{sk}$ | fraction of skeleton material (e.g., binder, carbon black, etc.) in electrode | % |
| $p_{sp}$ | separator porosity | % |
| $\tau_e$ | electrode tortuosity | – |
| $\tau_{sp}$ | separator tortuosity | – |
| $i$ | van't Hoff factor | – |
| $T$ | absolute temperature | K |
| $R$ | ideal gas constant = 8.3145 | J/mol/K |

# 2 Cell Data for Figs. 1 and 2

## 2.1 fte-CDI Cell Design and Testing

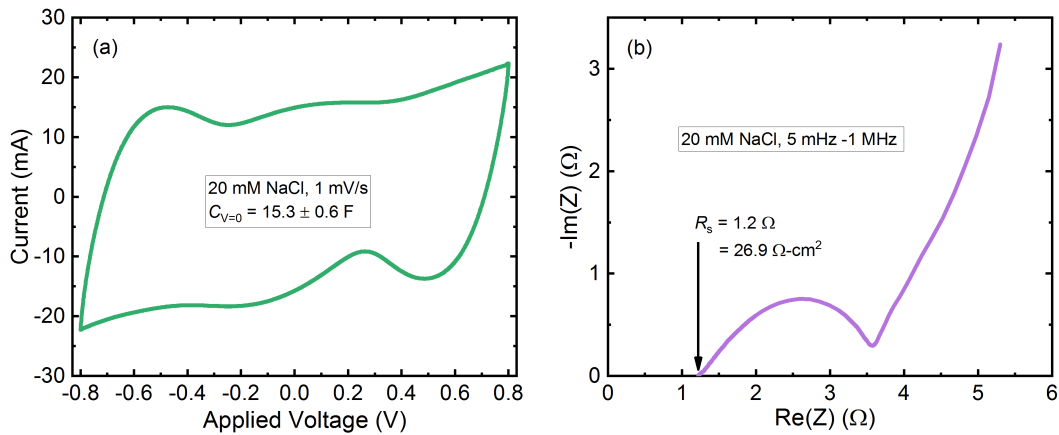

Figure S1: (a) CV and (b) EIS data for the cell in Figs. 3 and 4a,b of the main text and also detailed in the separation report below. A flow rate of 3.9 ml/min was used to avoid depletion effects.

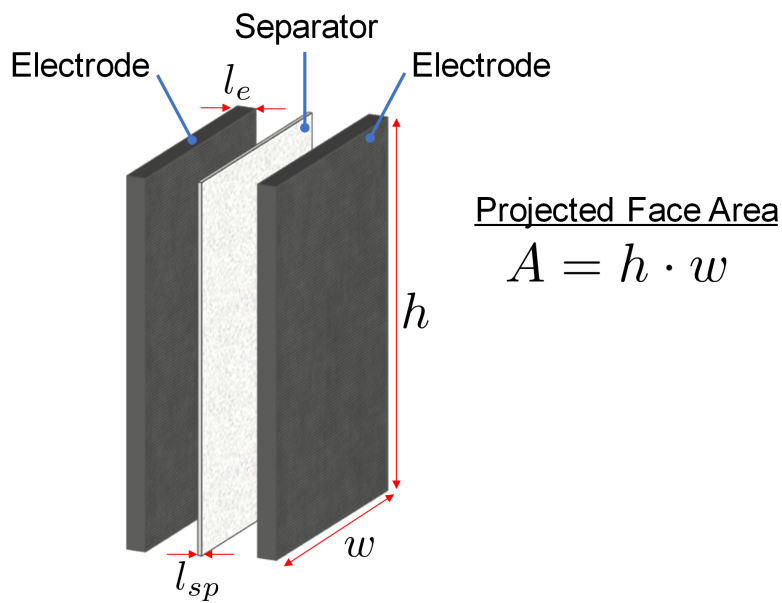

**Figure S2: An example exploded view diagram of the fte-CDI cell used herein. For this rectangular design, the projected face area** $A$ **is simply the width multiplied by the height.**

Fig. S1 is the corresponding CV and EIS data taken for the same device upon which Figs. 3 and 4a,b in the main text was generated. Fig. S2 is an example exploded view schematic illustrating the relevant cell dimensions for the fte-CDI device used herein. The flow rate for Fig. S1 (a) was 3.9 ml/min to avoid depletion effects. Below in Fig. S3 is the raw data underlying the performance data in Fig. 4a,b in the main text. In terms of obtaining this raw data, it is important to note how the separation was experimentally setup.[1] For example, if closed-loop recirculation was used or if the effluent was discarded; if the conductivity was measured with an in-line sensor or in the final reservoir; and if the feed water was purged with inert gas or simply equilibrated with air. For the data discussed herein we use the exact setup described in Ref. 2 (closed-loop circulation with a 2 L reservoir continuously purged with hydrated nitrogen). The results presented in Fig. 5 in the main text were measured with the total flow rates of 1-5 ml/min.

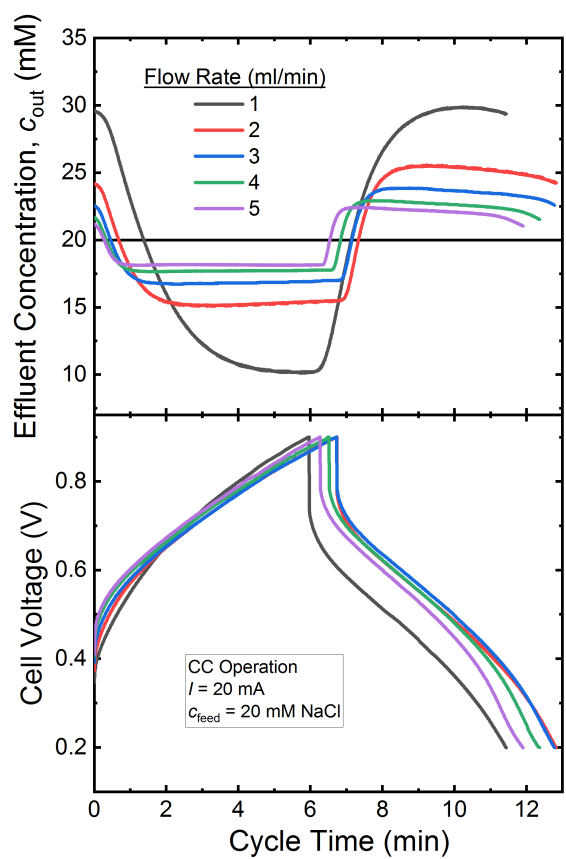

Figure S3: Experimental data of DSS concentration vs. time and voltage vs. time underlying the performance data in Fig. 4a,b. This cell is described in the separation report below and was measured in the manner described in Hawks et al.[2]

## 2.2 fb-MCDI Cell Design and Testing

fb-MCDI performance tests were carried out using a desalination stack described in detail by Zhao et al.[3] In total four cells with the total projected area of 135 cm$^2$ were used (33.75 cm$^2$ per cell). Each cell consists of two porous carbon electrodes (Materials & methods, PACMMTM 203, Irvine, CA, USA) and two ion exchange membranes, one selective for transport of cations and the other for anions (Neosepta, ASTOM Corporation, Tokyo, Japan). Each electrode mass is ∼0.55 gram and the thickness is ∼250 $\mu$m. The separator was a woven polymeric mesh with a thickness of ∼150 $\mu$m. In our MCDI experiments we used fixed feed water concentration of ∼20 mM which was pumped inside the MCDI stack from a 10 L vessel. The solution in the vessel was not flushed with nitrogen gas to remove dissolved oxygen as done in other studies.[4,5]

5

Typical experimental results of the effluent concentration and cell voltage obtained at different charge/discharge current densities are presented in Fig. S4. Based on this data it is possible to calculate performance indicators according to instructions given in the main text (Fig. S5). Experimental data are given for a cycle time (CT) of 2 and 6 min. Furthermore, the results presented in Fig. 5 in the main text were measured with the total flow rate of 7.5, 3.75, 2.5, and 1.875 mL/min per cell. The fb-MCDI cycles were defined by holding equal charging/discharging current, constant flow rate, preset discharge voltage (0 V), and varying the magnitude of charging/discharging current (0.74-1.85 mA/cm$^2$). The upper charging voltage was not controlled within the voltage window of 0.0-1.4 V. The discharge time was almost identical to the charging time; thus, the charging time is nearly equal to half the cycle time (CT).

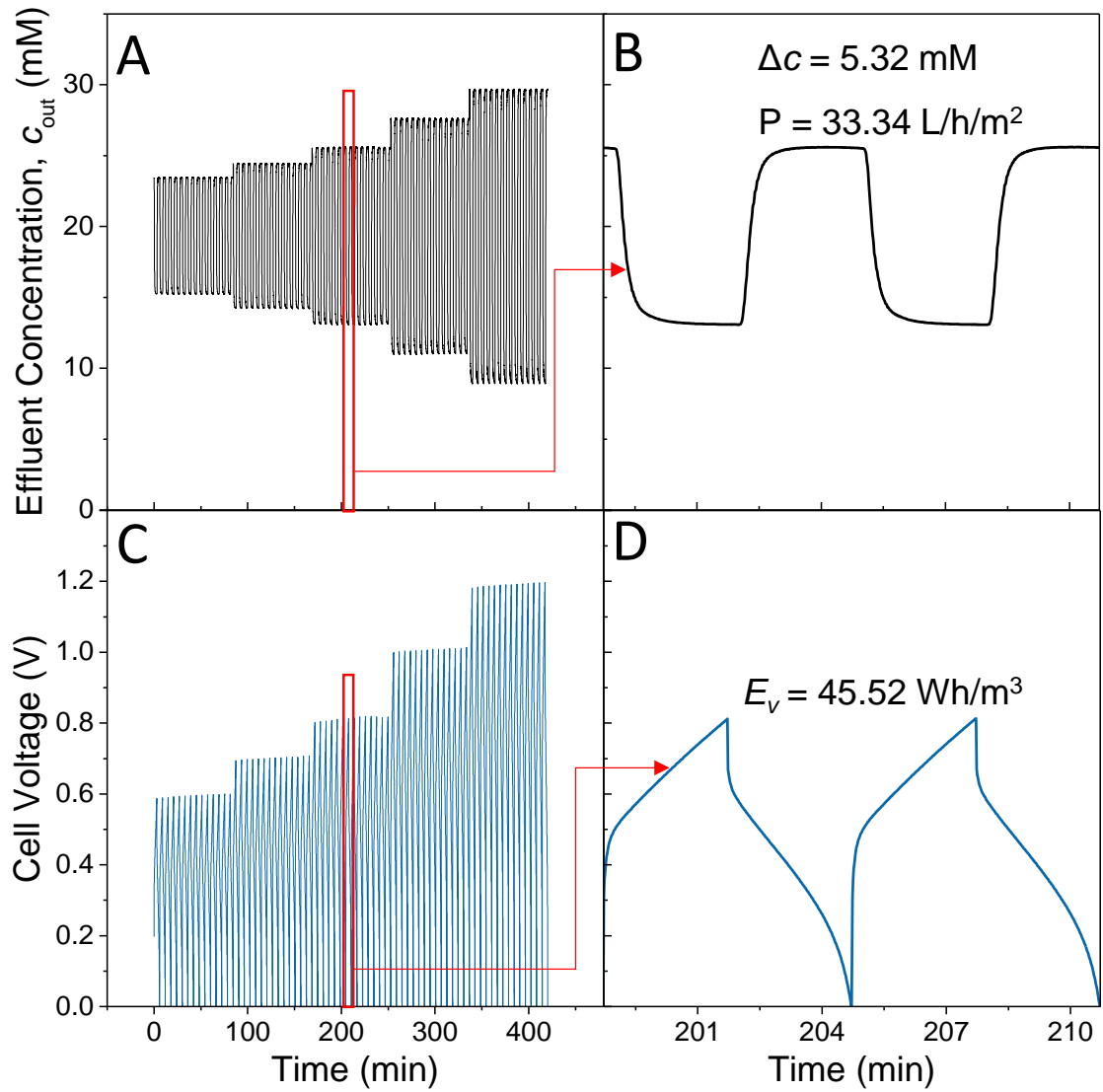

**Figure S4:** Experimental data underlying the performance data in Fig. 4c,d in the main text. The cycle time is 6 min and the flow rate is 3.75 ml/min per cell.

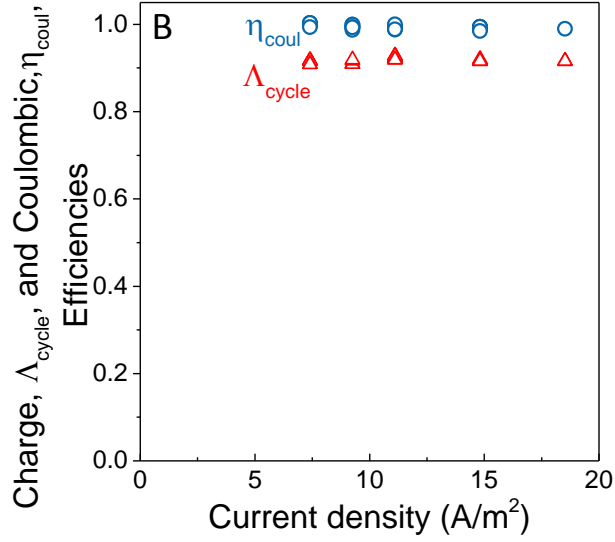

**Figure S5:** Experimental charge and Coulombic efficiency data underlying the performance data in Fig. 4c,d in the main text.

## 3 Thermodynamic Energy Consumption for a Separation

We use the equation for specific Gibbs free energy of separation as defined in Wang et al.,[6] which gives the thermodynamic minimum volumetric energy consumption ($E_{v,th}$) for a removal. In terms of the separation conditions discussed in this work, this equation is written as

$$\frac{E_{v,th}}{iRT} = \frac{c_{\text{feed}}}{WR} \ln \left[ \frac{c_{\text{feed}} - WR(c_{\text{feed}} - \langle \Delta c \rangle)}{c_{\text{feed}}(1-WR)} \right] - (c_{\text{feed}} - \langle \Delta c \rangle) \ln \left[ \frac{c_{\text{feed}} - WR(c_{\text{feed}} - \langle \Delta c \rangle)}{(c_{\text{feed}} - \langle \Delta c \rangle)(1-WR)} \right] \quad \text{(S1)}$$

where $i$ is the van't Hoff factor (here assumed to be 2), $R$ is the ideal gas constant (8.3145 J/mol/K), and $T$ is the absolute temperature (K) here assumed to be 294.15 K. Applying this equation to the separations in Fig. 4b in the main text and dividing by the experimentally measured $E_v$ gives the thermodynamic efficiency of the separation (Fig S6).[7] We refer to Hemmatifar et al.[7] for further details around the derivation of Eqn. (S1) from first principles, and a detailed discussion on the sources of irreversibilities and trade-offs in increasing the thermodynamic efficiency in practical CDI operation.

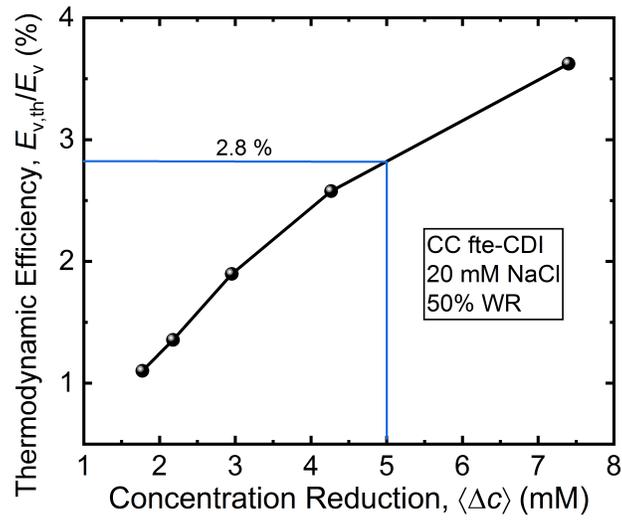

**Figure S6:** The thermodynamic efficiency for the energy consumption data presented in Fig. 4b in the main text. The 5 mM removal point has a productivity of 24 L/h/m$^2$.

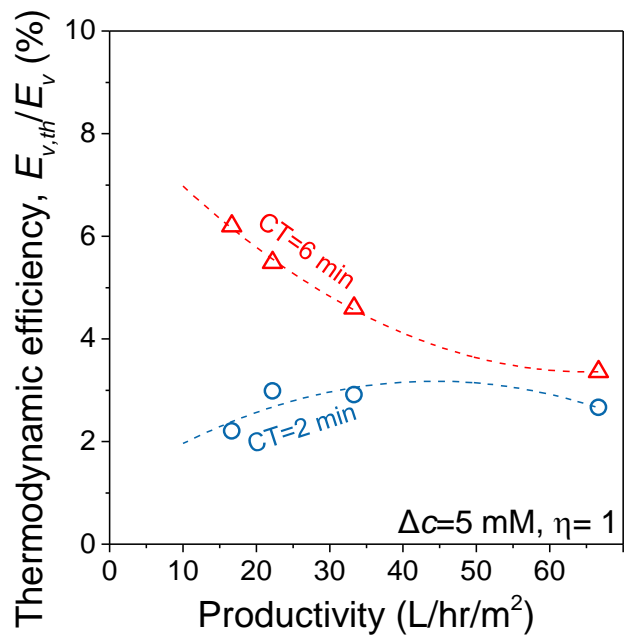

**Figure S7:** The thermodynamic efficiency for the energy consumption data presented in Fig. 5b in the main text.

9

# 4 Numerical Analysis of Experimental Cycle Data

Here we present a simple approach to obtaining the area of a function $f(t)$ above and below a given value $f_0$ between $t_0$ and $t_f$ (Fig. S8). This analysis can be readily applied to compute $N_{out}$, $E_{in}$, $E_{out}$, $V_d$, $V_c$, $q_{in}$, and $q_{out}$ discussed in the main text and defined above. One way of obtaining the areas of interest is as follows

$$|A| = \int_{t_0}^{t_f} |f(t) - f_0| \, dt \tag{S2}$$

$$\Delta A = \int_{t_0}^{t_f} f(t) - f_0 \, dt \tag{S3}$$

$$|A| = |A_+| + |A_-| \tag{S4}$$

$$\Delta A = |A_+| - |A_-| \tag{S5}$$

$$|A_+| = \frac{|A| + \Delta A}{2} \tag{S6}$$

$$|A_-| = \frac{|A| - \Delta A}{2} \tag{S7}$$

where $|A_+|$ and $|A_-|$ are the magnitudes of the area of $f(t)$ above and below $f_0$ between $t_0$ and $t_f$, respectively.

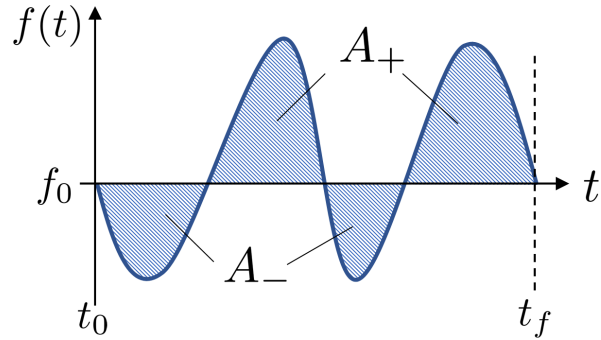

Figure S8: **An illustration of a function with areas $A_+$ and $A_-$ of interest.**

This formulation is useful because only integrals over the entire DSS cycle need to be computed (Eqns. (S3) and (S2)). In general, appropriate manipulation of the sign of the integrand can yield the areas of interest from this approach (see cycle calculator in the accompanying spreadsheet). Another

numerical procedure of interest (also in the accompanying spreadsheet) is converting conductivity data to salt concentration. This task is outlined in detail in the SI of Ref. 2 below.

## 5  Measuring Capacitance and PZC with CV

Here we provide example CV analysis for examining material properties relevant to CDI operation. Fig. S9 shows an example 2-electrode cell measured over several NaCl concentrations from CDI relevant (20 mM NaCl) to energy-storage relevant (2 M NaCl). The measurement in Fig. S9 was taken using two identical activated carbon electrodes separated by a polyester mesh (McMaster 9218T73) and contacted with a custom titanium 4-wire point contact clamp in a stirred 0.5-1 L beaker. The material used to obtain Fig. S9 was hierarchical carbon aerogel monolith (HCAM) derived from carbonized and activated resorcinol-formaldehyde aerogel.[8–10] The dimensions of the electrodes were 2×2.5 cm with an average thickness of 365 $\mu$m and mass of 0.098 g per electrode, which corresponds to a density of 0.54 g/cm$^3$. Assuming a solid carbon density of 1.95 g/cm$^3$ implies a total porosity ($p_{mi} + p_{mA}$) of 72.5%.

Fig. S9 (a) shows DSS CV curves for various NaCl concentrations, including the CDI relevant case of 20 mM. The PZC is indicated on the positive-voltage charging side of the curve, but because it is an identical 2-electrode system, the PZC minima is also witnessed on the discharge and negative-voltage half of the measurement. Fig. S9 (b) shows the corresponding capacitance for each concentration taken from the average discharge PZC local minima ($C_{PZC}$) or at $V = 0$ ($C_{V=0}$). The PZC capacitance at CDI relevant concentrations (20 mM) varies strongly with ionic strength and is 67% of that at energy-storage relevant concentrations. Moreover, there is a significant difference between the capacitance evaluated at zero applied voltage and at the PZC, implying that it is important to take into account how the capacitance was evaluated when comparing different materials at CDI-relevant concentrations.

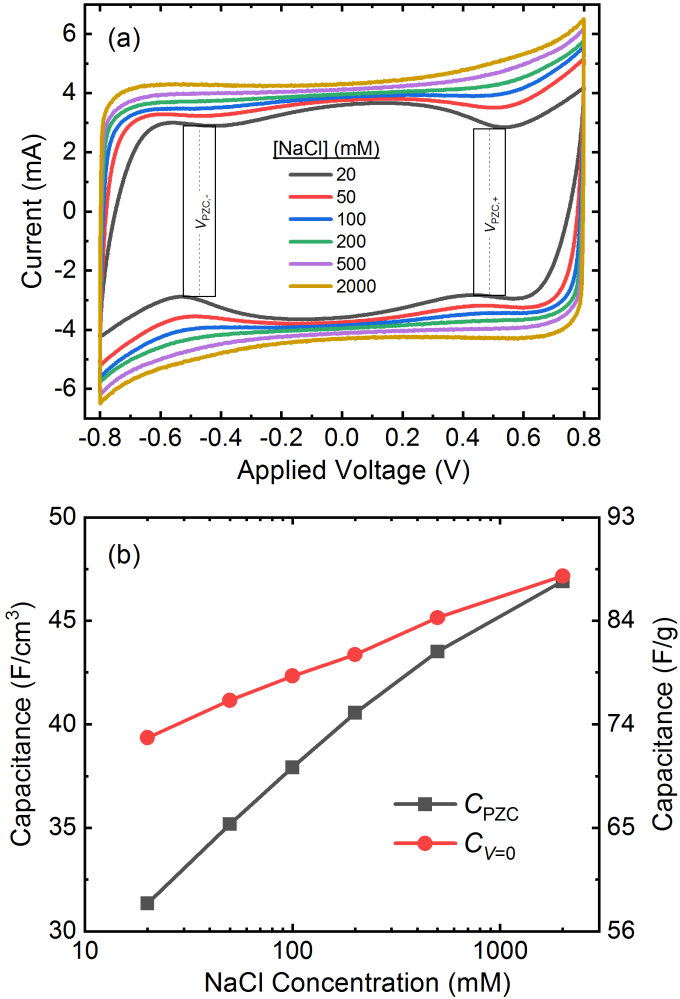

**Figure S9:** (a) 1 mV/s CV curves for a 2×2.5 cm two electrode cell contacted with a 4-point probe in a stirred 0.5-1 L beaker containing NaCl and deionized water. (b) Resulting HCAM material capacitance values from the curves in (a) showing that the cell capacity at PZC in 20 mM is 67% of that at 2 M. The PZC is located approximately symmetrically at ±0.48 V.

## 6 Measuring Series Resistance with EIS

Here we provide examples for how to measure $R_s$, $R_{EER}$, and $R_{sp}$ via EIS. Fig. S10 (a) shows EIS spectra taken from 700 kHz to 40 mHz on a 2×2 cm stacked HCAM assembly in 20 mM (2.37 mS/cm) NaCl solution with a 4-wire point-probe connection and 1 mm thick electrodes at zero applied bias. Here we define $R_s$ in terms of the EIS Nyquist plot as the high frequency value of Re(Z) at which -Im(Z) tends to zero. Fig. S10 (a) clearly reveals that increasing the separator thickness by stacking multiple sheets simply shifts the Nyquist plot to higher Re(Z) values, leaving the overall

shape intact. Thus, as expected, plotting $R_s$ as a function of number of separators stacked together in Fig. S10 (b) gives a straight line that reveals $R_{EER}$ and $R_s$. If $R_s$ was captured by a different region of the Nyquist plot, then the overall shape would be expected to change significantly, which is not the case in Fig. S10 (a). To further illustrate how EIS can be used to measure $R_s$, we also plot in Fig. S11 (a) the resulting EIS Nyquist data for the concentration series later discussed in Fig. S9 at zero applied bias. Fig. S11 (a) also shows a complete shift in the Nyquist data to lower Re(Z) with higher NaCl concentration. While the total shape of the Nyquist plot provides information about all the resistances of the system,[11] the intercept where -Im(Z) equals zero reveals $R_s$ specifically. Since $R_s$ includes the separator resistance (Eqn. (24) in the main text), it is expected to scale linearly with solution resistivity (approximately inversely with concentration). Fig. S11 (b) shows that this is indeed the case. Thus, a simple EIS concentration series can be measured to reveal the underlying components of the series resistance. This method is readily applicable to fully assembled cells, which can help identify a significant portion of the total resistance breakdown of the device.[5,11–13]

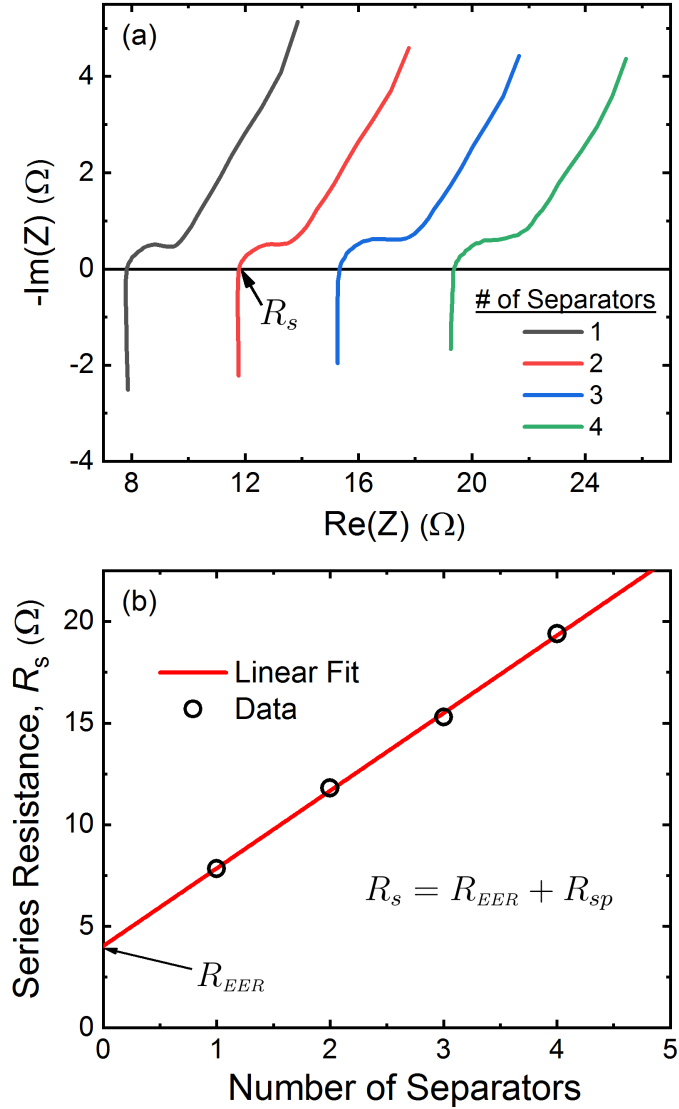

Figure S10: (a) EIS data taken from 700 kHz to 40 mHz on a 2×2 cm stacked HCAM assembly in 20 mM (2.3 mS/cm) NaCl solution with a 4-wire point-probe connection at zero applied bias. (b) Corresponding plots of $R_s$ from (a) as a function of number of 125 $\mu$m coffee filter paper separators stacked on top of each other. The $R_s$ intercepts occur around 40 kHz, and the slope of the line corresponds to 3.82 $\Omega$/separator for this particular arrangement.

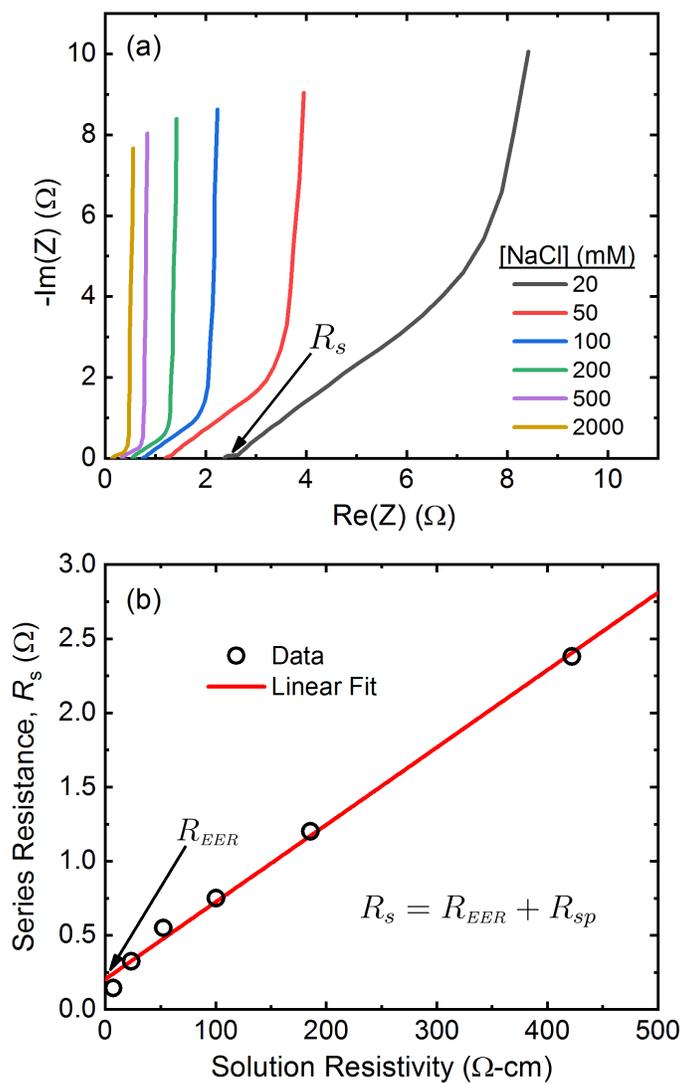

**Figure S11:** (a) EIS data taken from 700 kHz to 5 mHz at zero applied bias on the same assembly as in Fig. S9. (b) Corresponding plots of $R_s$ from (a) as a function of solution resistivity. The separator was a 90 $\mu$m polyester plastic mesh (McMaster 9218T73). The frequencies corresponding to the $R_s$ intercepts range from approximately 500 Hz (2 M NaCl) to 125 kHz (20 mM NaCl), and the slope of the line for this particular arrangement was 0.521 m$^{-1}$.

15

# 7 Separation Report

While such experimental details may seem trivial, it is critical to also report the basic cell dimensions and material characteristics that are behind the separation. The cell face area $A$ and thickness $l_e$ are important values for normalizing and understanding performance. For example, the electrode volume is used for normalization of capacitance is $A \cdot l_e$, and the face area of the cell is used for normalizing productivity. If membranes are used, the face area and number of membranes used ($n_m$) are especially important for estimating device cost, as membranes are typically \$200/m$^2$ or more. Additionally, basic electrode characteristics like the mass per electrode (m) and density ($\rho_e$) are useful for categorizing material performance. Finally, porosity estimations of the separator and electrode material (microporosity $p_{mi}$, macroporosity $p_{mA}$, "skeleton" porosity $p_{sk}$, etc.)[5,14] are critical for understanding cell resistance and fluid volume.[2,12]

Thus, to ensure that critical characteristics are reported, we propose adding a separation report to the SI of a manuscript similar to that shown below, which corresponds to the data in Fig. 2 in the main text. For convenience, we also add an excel spreadsheet to the electronic SI that can be downloaded along with this document. If parts of this report are not measured or known, a partially filled in report is still highly valuable.

**Table S2: CDI Separation Report for the fte-CDI Cell in Fig. 4a,b at 5 mM Removal**

| Separation Conditions | | |
|---|---|---|
| Parameter | Value | Notes |
| ⟨Δc⟩ (mM) | 5 | Average concentration reduction of the DSS cycle. |
| $c_{feed}$ (mM) | 20 | Feed water salt concentration. |
| WR (%) | 50 | Water recovery. |
| κ (mS/cm) | 2.8 | Measured feed water electrical conductivity. |
| Salt | NaCl | Feed water composition. |
| Operation Mode | CC | Constant current, constant flow rate, 0.2-0.9 V voltage window |

| Performance Metrics | | |
|---|---|---|
| Parameter | Value/Figure | Notes |
| P (L/h/m$^2$) | Fig. 5 | Productivity. Volume of diluate per total cycle time per projected face area. |
| $E_v$, η = 1 (Wh/m$^3$) | Fig. 5 | Volumetric energy consumption with full energy recovery and reuse. |
| $E_v$, η = 0.5 (Wh/m$^3$) | -- | Volumetric energy consumption with 50% energy recovery and reuse. |
| $E_v$, η = 0 (Wh/m$^3$) | Fig. 5 | Volumetric energy consumption with no energy recovery and reuse. |

| Performance Indicators | | |
|---|---|---|
| Parameter | Value | Notes |
| Eq-SAC (mg/g or mmol/cm$^3$) | -- | Not measured. Equilibrium salt adsorption capacity. |
| C (F) | 16.3 ± 2.2 | Cell differential capacitance in 20 mM NaCl. Measured here from the discharge during CC operation. |
| C (F/g) | 45.3 ± 6.1 | Calculated from cell capacitance and electrode mass. |
| C (F/cm$^3$) | 25.3 ± 3.4 | Calculated from cell capacitance and electrode volume. |
| $Λ_{cycle}$ (%) | 65.4 | DSS total cycle charge efficiency. |
| $η_{coul}$ (%) | 90.6 | DSS cycle Coulombic efficiency. |
| $R_s$ (Ω-cm$^2$) | 26.9 | Measured from EIS in 20 mM NaCl, see Fig. S1 |

| Cell Characteristics | | |
|---|---|---|
| Parameter | Value | Notes |
| n | 1 | Number of cells (electrode pairs) used in the separation. |
| $n_m$ | 0 | Number membranes used per cell. |
| A (cm$^2$) | 22.4 | Projected face area per CDI cell. |
| $l_e$ (μm) | 575 | Electrode thickness. |
| $l_{sp}$ (μm) | 125 | Separator thickness. |
| m (g) | 1.44 | Total mass of electrodes. |
| $ρ_e$ (g/cm$^3$) | 0.56 | Electrode mass density. |
| $σ_e$ (S/cm) | 22 ± 8 | Electrode material electrical conductivity. Here measured with the Van der Pauw method. |
| $p_{mi}$ (%) | 20 | Electrode microporosity. Estimated based on typical literature values. |
| $p_{sk}$ (%) | 0 | Volume fraction of "skeleton" material (e.g., binder, carbon black, etc.). |
| $p_{mA}$ (%) | 51 | Electrode macroporosity. Calculated assuming $ρ_{carbon}$ = 1.95 g/cm$^3$, $p_{mi}$ = 20 %, and $p_{mA}$ = 1 - $ρ_e/ρ_{carbon}$ - $p_{mi}$. |
| $τ_e$ | 1.4 | Electrode tortuosity. Estimated from Bruggeman relation $τ_e = 1/\sqrt{p_{mA}}$. |
| $p_{sp}$ (%) | 68 | Separator porosity. Estimated from water intrusion. |
| $τ_{sp}$ | 2 | Separator tortuosity. Estimated from EIS measurements of stacked separators. |

**Table S3: CDI Separation Report for fb-MCDI Cell in Fig. 4c,d at 5 mM Removal**

| Separation Conditions | | |
|---|---|---|
| Parameter | Value | Notes |
| $\langle\Delta c\rangle$ (mM) | 5 | CC (discharge until zero voltage), with different current until $\langle\Delta c\rangle$ is below and above 5 mM, and then interpolate |
| $c_{feed}$ (mM) | 19.2 | |
| WR (%) | 50 | |
| $\kappa$ (mS/cm) | 2.24 | |
| Salt | NaCl | |
| Operation Mode | CC | Constant current |

| Performance Metrics | | |
|---|---|---|
| Parameter | Value | Notes |
| P (L/h/m$^2$) | Fig. 5 | See Fig. 5 in the main text. |
| $E_v$, $\eta$ = 1 (Wh/m$^3$) | Fig. 5 | See Fig. 5 in the main text. |
| $E_v$, $\eta$ = 0.5 (Wh/m$^3$) | -- | |
| $E_v$, $\eta$ = 0 (Wh/m$^3$) | Fig. 5 | See Fig. 5 in the main text. |

| Performance Indicators | | |
|---|---|---|
| Parameter | Value | Notes |
| Eq-SAC (mg/g or mmol/cm$^3$) | -- | |
| C (F) | 78 | Estimated from the slope of cell voltage vs. time curve for a CC charge |
| C (F/g) | 35 | Using $C$ above and $m$ below |
| C (F/cm$^3$) | 23 | Using F/g above and $p_e$ below |
| $\Lambda_{cycle}$ (%) | 92 | |
| $\eta_{coul}$ (%) | 99 | |
| $R_s$ ($\Omega$-cm$^2$) | 135 | Estimated from the initial voltage jump upon CC charging |

| Cell Characteristics | | |
|---|---|---|
| Parameter | Value | Notes |
| n | 4 | Number of cells (electrode pairs) used in the separation. |
| $n_m$ | 8 | Per cell there is one cation and one anion exchange membrane |
| A (cm$^2$) | 33.75 | Projected face area per CDI cell. |
| $l_e$ ($\mu$m) | 250 | Electrode thickness. |
| $l_{sp}$ ($\mu$m) | 150 | Separator thickness. |
| m (g) | 4.42 | Total mass of electrodes. |
| $\rho_e$ (g/cm$^3$) | 0.65 | Electrode mass density. |
| $\sigma_e$ (S/cm) | - | |
| $p_{mi}$ (%) | - | |
| $p_{sk}$ (%) | - | |
| $p_{mA}$ (%) | - | |
| $\tau_e$ | - | |
| $p_{sp}$ (%) | - | |
| $\tau_{sp}$ | - | |


\end{document}